\documentclass[journal,twocolumn]{IEEEtran}
\usepackage{bbm}
\usepackage{algorithm}
\usepackage{algorithmic}
\usepackage[font=small]{caption}
\usepackage[labelsep=period]{caption}
\def\delequal{\stackrel{\triangle}{=}}
\usepackage{float}
\usepackage{stackengine}
\usepackage[latin1]{inputenc}
\usepackage{graphicx}
\usepackage[cmex10]{amsmath}
\usepackage{cite}
\usepackage{amsfonts}
\usepackage{amsmath}
\usepackage{amsthm}
\usepackage{mathrsfs,dsfont}
\usepackage{amssymb,mathrsfs}
\usepackage[mathscr]{euscript}
\usepackage{epstopdf}

\usepackage{color}
\usepackage{pgfplots}
\pgfplotsset{compat=newest}
\usetikzlibrary{plotmarks}
\usepackage{grffile}
\usepackage{amsmath}
\usepackage{caption}
\usepackage{subcaption}
\usepackage{balance}
\newcommand{\argmax}{\arg\!\max}
\IEEEoverridecommandlockouts

\usepackage{soul}
\begin{document}


\author{\IEEEauthorblockN{
Atefeh Hajijamali Arani\IEEEauthorrefmark{1},
Peng Hu\IEEEauthorrefmark{1}\IEEEauthorrefmark{2}, and
Yeying Zhu\IEEEauthorrefmark{1}}

\IEEEauthorblockA{\IEEEauthorrefmark{1}University of Waterloo, ON N2L 3G1, Canada}

\IEEEauthorblockA{\IEEEauthorrefmark{2}National Research Council of Canada, Waterloo, ON N2L 3G1, Canada}

\thanks{Atefeh Hajijamali Arani and Yeying Zhu are with the Dept. of Statistics and Actuarial Science, University of Waterloo, ON N2L 3G1, Canada (e-mail: ahajijam@uwaterloo.ca; yeying.zhu@uwaterloo.ca).}
\thanks{Peng Hu is with Digital Technologies Research Center, National Research Council Canada, Waterloo, ON N2L 3G1, Canada, and with the Cheriton School of Computer Science, University of Waterloo, ON N2L 3G1, Canada. (Corresponding author email: Peng.Hu@nrc-cnrc.gc.ca)}
}


\title{HAPS-UAV-Enabled Heterogeneous Networks: \\A Deep Reinforcement Learning Approach}
\maketitle

%
\IEEEpeerreviewmaketitle

\begin{abstract}
The integrated use of non-terrestrial network (NTN) entities such as the high-altitude platform station (HAPS) and low-altitude platform station (LAPS) has become essential elements in the space-air-ground integrated networks (SAGINs). However, the complexity, mobility, and heterogeneity of NTN entities and resources present various challenges from system design to deployment. This paper proposes a novel approach to designing a heterogeneous network consisting of HAPSs and unmanned aerial vehicles (UAVs) being LAPS entities. Our approach involves jointly optimizing the three-dimensional trajectory and channel allocation for aerial base stations, with a focus on ensuring fairness and the provision of quality of service (QoS) to ground users. Furthermore, we consider the load on base stations and incorporate this information into the optimization problem. The proposed approach utilizes a combination of deep reinforcement learning and fixed-point iteration techniques to determine the UAV locations and channel allocation strategies. Simulation results reveal that our proposed deep learning-based approach significantly outperforms learning-based and conventional benchmark models. 

\end{abstract}

\begin{IEEEkeywords}
Deep reinforcement learning, high-altitude platform station, resource allocation, fairness, unmanned aerial vehicles, non-terrestrial networks
\end{IEEEkeywords}

\section{Introduction}
Recently, the integrated use of non-terrestrial network (NTN) entities such as high-altitude platform stations (HAPSs) and low-altitude platform stations (LAPSs) has become essential elements in the space-air-ground integrated networks (SAGINs). These entities can complement the space segments to provide high-quality network access to global users. For example, the use of HAPSs together with unmanned aerial vehicles (UAVs) being LAPS can be well suited for meeting capacity and coverage demands, such as temporary events-driven coverage, greenfield coverage, terrestrial backhaul, and white spot reduction \cite{GSMA2022}, with the capability of keeping the round-trip-time latency down to within 10 ms. However, the great promises of such HAPSs and LAPSs in a SAGIN come with challenges. One challenge in an important and generic scenario is fairness assurance in the overall quality of experience (QoE) for ground users. The heterogeneous NTN entities, resources, and dynamics of UAV trajectories add much complexity to the system modeling and solutions. 

Most recent works have proposed to address UAVs and HAPS separately. For UAVs, the recent results are focused on trajectory and resource management within a UAV network where multiple UAVs are employed. The optimization of UAV trajectory for quality of service (QoS) performance and coverage is discussed in \cite{Hao_WCSP20, Guo_IWCMC19}. Deep Q-network (DQN) \cite{Wang_JOCIN20, Liu20_TVT} and reinforcement learning methods \cite{atefeh_iot22}  have been applied to UAV trajectory optimization, while deep learning methods \cite{Tang20_highmob, Tang_JSAC22, atefeh_icc2021} for optimal resource management have been proposed. HAPS has been studied as a standalone system providing uplink and downlink to the ground users \cite{Yuan22} and as part of a SAGIN system \cite{Jia21_JSAC, Ahmadinejad22}. 

In order to address the dynamic nature of an integrated system consisting of HAPS and LAPS entities, Q-learning is considered an effective technique for solving an optimal solution in system modeling. However, it poses limitations when the mobility of ground users and UAVs is considered. Furthermore, it must deal with the exponential growth of states and actions when exploring an optimal solution in a high-dimensional space. A new approach is needed to solve the theoretical limits while meeting generic QoE or QoS requirements in an integrated system setting. In the context of a HAPS-UAV-enabled heterogeneous network, such an approach needs to be applied to the fundamental challenge in resource allocation and UAV trajectory planning, considering practical deployment configurations. This challenge has hardly been well addressed in the current works.

This work proposes a deep reinforcement learning-based algorithm for aerial base stations (ABSs) to provide network services in a highly dynamic environment where the mobility of ground users and UAVs presents a challenge for conventional reinforcement learning techniques such as Q-learning. This is due to the potential for failure caused by the curse of dimensionality. To address this issue, the proposed algorithm uses neural networks to approximate Q-value functions, allowing the UAVs to operate autonomously and intelligently adapt to rapidly changing conditions. In particular,
we make the following contributions:

\begin{itemize}
\item We construct a high dynamic scenario of an aerial heterogeneous
 network in a diverse environment, considering both HAPSs and UAVs while taking into account user mobility. 

\item We employ the deep reinforcement learning algorithm to intelligently optimize the trajectory and transmit channel of UAVs. 

\item   Our proposed solution considers  loads of ABSs and incorporates them into the trajectory design and resource allocation process which determines the average resource utilization at ABSs and the system's ability to provide sufficient QoS to  users. Furthermore, we optimize fairness among users in the system. 

\item We compare the performance in terms of fairness, rate, and outage  between the proposed and
reinforcement learning-based benchmarks.
\end{itemize}

The remainder of the paper is structured as follows. Section II overviews the related work. Section III presents the system model and problem statement. Section IV discusses the Q-learning and our proposed DQN-based scheme for a joint resource management and trajectory design. Section V evaluates the proposed scheme in comparison with the typical algorithms and variations. 

\section{Related Work}
 In recent years, the integration of HAPSs and UAVs into communication networks has gained significant attention as a promising solution for extending wireless coverage and providing access to remote areas. 
 HAPS provides a high-altitude persistent coverage that can reduce the number of cell towers required, resulting in lower capital and operational costs. Furthermore, the mobility of UAVs allows for dynamic deployment in areas with high user density, thereby improving the overall network capacity.  The use of multiple HAPSs and UAVs in a network can also provide improved reliability.
 Most of the works  focus on optimizing the trajectory of UAVs to enhance network performance and coverage. These studies have proposed various trajectory design algorithms based on non-learning and learning algorithms, with the aim of maximizing the network's coverage area and enhancing user throughput. In \cite{Hao_WCSP20}, a trajectory design algorithm based on deep reinforcement learning for a single UAV is proposed. The solution aims at maximizing the uplink sum rate of users. To maximize the spectral efficiency of a network composed of a ground base station (BS) and UAVs, a deep reinforcement algorithm to optimize the locations of UAVs is developed in \cite{Guo_IWCMC19}. Moreover, it is assumed that users have different QoS. In \cite{Wang_JOCIN20}, a UAV is employed for emergency communication support for users. The objective is to maximize the number of served users  and  uplink data rate by optimizing the UAV trajectory and transmission power of users. A DQN-based algorithm is proposed to solve the UAV trajectory problem. Additionally, a successive convex approximation-based algorithm is proposed for power control at the level of  users, based on the optimized UAV trajectory. To optimize the trajectory of a single UAV for mobile edge computing, a double deep Q-network algorithm is proposed in  \cite{Liu20_TVT}. The authors in \cite{Tang20_highmob} utilize a deep learning algorithm for  dynamically allocating radio resources for uplink and downlink.  In \cite{Tang_JSAC22}, the authors propose a reinforcement learning approach to address the challenge of traffic offloading in an aerial network. The proposed solution employs a double Q-learning algorithm with an improved delay-sensitive replay memory mechanism to train the nodes to make intelligent offloading decisions based on both local and neighboring historical information. Additionally, they utilize a joint information collection technique and an offline training mechanism to further enhance the efficiency of the algorithm. In \cite{atefeh_iot22}, the authors propose an energy-efficient UAV path planning based on reinforcement learning and satisfaction algorithms.   To maximize the throughput of an aerial network, learning-based mechanisms are implemented in \cite{atefeh_icc2021, APY_Access2021}. In \cite{APY2022}, the learning algorithms are surveyed in UAV-assisted SAGINs. However, the existing studies are restricted to only UAV networks and do not consider HAPS. 

 On the other hand,  the integration of HAPSs in UAV networks can  enhance the capabilities of aerial networks, providing a cost-effective solution to meet the increasing demands for high-speed and reliable communication.  In \cite {Cao21}, the authors propose a transmission scheme that combines the HAPS and the ground-to-space transmission to improve terrestrial communication and reduce transmission power. They develop a transmission control strategy, where ground users can switch between the two transmission schemes with a probability, which is determined to maximize overall throughput.
In \cite{Swaminathan21}, a solution is developed  to improve the reliability of uplink  communications by fusing free-space optics (FSO) and radio frequency (RF) technologies. The proposed solution utilizes a HAPS, as a relay station. Furthermore, two system models, single-hop and SAGIN-based dual-hop are investigated for uplink communication with hybrid FSO/RF links. 
The impact of HAPS deployments on terrestrial networks is investigated in \cite{Yuan22}. It analyzes both co-channel and adjacent channel deployment scenarios with both unsynchronized and synchronized time-division duplexing (TDD). The results indicate that the synchronized TDD scenario requires a smaller inter-system distance than the unsynchronized case for the co-channel case. In the adjacent channel case, the interference-to-noise ratio  is always below a certain threshold  for both unsynchronized and synchronized scenarios. Furthermore, results for full buffer and bursty traffic models under different traffic loads are considered.
In \cite{Jia21_JSAC}, low Earth orbit (LEO) satellites and  HAPSs are used  to provide  access and data backhaul to remote area users which aims at maximizing the revenue of LEO satellites. The problem is formulated as a  mixed integer nonlinear programming. To solve the problem, matching algorithms are proposed. In \cite{Ahmadinejad22}, the authors consider a system composed of a HAPS and a set of UAVs, in which  the locations of all the ABSs are fixed. To solve the problem of power and sub-carrier allocation, a heuristic greedy algorithm is used. However, the aforementioned work focused on HAPSs does not take into account the fairness issue in the system and most studies consider statistics scenarios for users in the system. 

\section{System Model and Problem Statement}
\subsection{System Model}
\begin{figure}[t]
\centering\resizebox{3.5in}{2.6in} {\includegraphics
{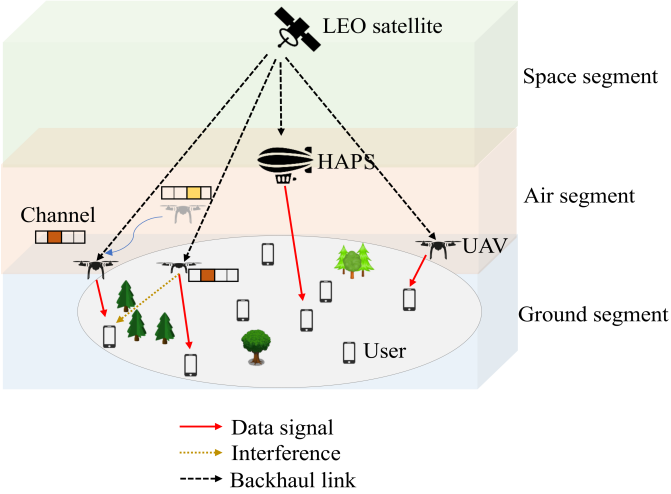}}
\caption{An illustration of the system model.} \label{layout}
\end{figure}
In this section, we present the system formulation and the problem statement. 
As depicted in {Fig. \ref{layout}}, the considered HAPS-UAV-enabled heterogeneous network is heterogeneously constructed with HAPSs and low altitude platforms (LAPs) or UAVs.  
We consider the downlink transmission of the system  composed of a set of UAVs $\mathcal U$ and a set of HAPSs $\mathcal M$ as ABSs.  Let $\mathcal B = \mathcal U \cup  \mathcal M$ denote the set of total ABSs in the system.  Furthermore, we assume that LEO satellites provide backhaul connectivity for the 
 ABSs. 
The set of total users and the set of users associated with ABS $b\in\mathcal B$ at time instant $t$ are represented by $\mathcal K$ and $\mathcal K_b(t)\in\mathcal K$, respectively. 
The  three-dimensional ($3$D)  location of ABS $b$ is denoted by  $\boldsymbol z_b^{\mathrm {ABS}}(t) = (x_b(t), y_b(t), h_b(t))$, where $(x_b(t), y_b(t))$ and $h_b(t)$ are the horizontal coordinate and the altitude of ABS $b$ at time instant $t$, respectively. Generally, discrete-time  sampling is adopted to update the system configuration. We consider a discrete-time setting $\mathcal N =\{0, 1, 2, \dots, N\}$.
We assume that the HAPSs are fixed and the UAVs fly at a fixed speed $v_{\mathrm {U}}$. Therefore, the location of UAV $u\in \mathcal U$ is updated as follows:
\begin{equation}\label{eq_UP_LO}
\boldsymbol z_u^{\mathrm {ABS}}(t+1) = \boldsymbol z_u^{\mathrm {ABS}}(t) + v_{\mathrm {U}}(t) T_{\mathrm s}, 
\end{equation}
where $T_{\mathrm s}$ and $\boldsymbol z_u^{\mathrm {ABS}}(t)$ are the duration of each time slot and the location of UAV $u$ at time instant $t$, respectively.

\subsection{User Mobility Model} \label{ue_mob_sec}
We assume that the users move according to a random walk  mobility model \cite{mobility2002survey}.  Let $\boldsymbol z_k^{\mathrm {UE}}(t) = (x_k(t), y_k(t), h_k)$ denote the coordinate of user $k\in\mathcal K$ at time instant $t\in \mathcal N$, where $(x_k(t), y_k(t))$ and $h_k$ are the horizontal coordinate and the height of user $k$ at time instant $t$, respectively. 
Obviously, the heights of the users are fixed.
In this model, the users change their speeds and movement directions with zero pause time at each time slot. 
At each time, the speed of user $k$, $v_k(t)$, is  randomly determined from the predefined ranges  $[v^{\mathrm{min}}_\mathrm{UE}, v^{\mathrm{max}}_\mathrm{UE}]$ following a uniform distribution, where $v^{\mathrm{min}}_\mathrm{UE}$ and $v^{\mathrm{max}}_\mathrm{UE}$ denote the minimum  and  maximum speed of the users, respectively. Furthermore, the  movement direction for user $k$, $\phi_k(t)$, is randomly chosen from the ranges $[0, 2\pi]$ according to a uniform distribution. Therefore, for each user $k\in\mathcal K$, the velocity vector is $[v_k(t)\cos\phi_k(t), v_k(t)\sin\phi_k(t)]$.

\subsection{Radio Propagation and Signal Quality}

We assume that at each time slot, the network topology is quasi-static, and the channel state information is constant. 
We adopt the  International Telecommunications Union (ITU) path loss model between the users and the ABSs. The path loss model between  HAPS $m\in\mathcal M$ and user $k\in\mathcal K$ includes the free space path loss (FSPL) model which can be expressed as \cite{Shibata22_acess}

\begin{equation}\label{eq_UP_LO}
L_{m,k}(t) = 32.44  + 20 \log_{10} f_{\mathrm{HAPS}} + 20 \log_{10} d_{m, k}(t)\  \ [\mathrm{dB}],
\end{equation}
where $f_{\mathrm{HAPS}}$ and $d_{m, k}(t)$ are the HAPS' operating frequency in Mega Hertz (MHz)  and the distance in kilometers between user $k$ and  HAPS $m$ at time $t$, respectively. 

To model a channel between  user  $k$ and  UAV $u$, we consider the  model described in \eqref{prob_LoS} which includes   line-of-sight (LoS)  and non-LoS components.  
The probability of   having a LoS link between user $k$ and UAV $u$ depends on the environmental characteristics and it  can be written  as ~\cite{atefeh_iot22} 

\begin{equation}\label{prob_LoS}
 \mathrm{pr}_{u,k}^{\mathrm {LoS}}(t) =  \prod_{n=0}^J\!\Bigg[1-\exp\! \Bigg(\!-\frac{{\Big[\!h_u(t)-\frac{(n+\frac{1}{2})(h_u(t)-h_k\!)}{J+1}\!\Big]}^2}{2 \xi^2}\!\Bigg)\!\Bigg],
\end{equation}
 where $J = {\lfloor\frac{r_{u,k}(t)  \sqrt{\alpha  \beta}}{1000}-1 \rfloor}$, and  $\alpha$, $ \beta$ and $\xi$ 
represent  statistical environment-dependent parameters \cite[Table 1]{antenna2008}. 
Here, parameter  $\alpha$ represents the ratio of land area  covered by buildings to total land area, $\beta$ denotes the mean number of buildings per unit area, and $\xi$ is the distribution of building height. This blockage model  can be used for air-to-ground transmissions with any transmitter/receiver heights   and for a broad  spectrum range \cite{hamed_ahmadi2020}. 
Here, $r_{u,k}(t) = \sqrt{(x_u(t) - x_k(t))^2 + (y_u(t) - y_k(t))^2}$ denotes the  horizontal distance between UAV $u\in\mathcal U$  and  user $k\in\mathcal K$ at time $t$.
Therefore,  the probability of having a non-LoS link at time $t$ can be determined as   $\mathrm{pr}_{u,k}^{\mathrm {NLoS}}(t) = 1- \mathrm{pr}_{u,k}^{\mathrm {LoS}}$(t).

Let $d_{u,k}(t) = \sqrt{r_{u,k}^2(t) + (h_u(t) - h_k)^2}$ be the $3$D distance between UAV $u$ and user $k$ at time $t$. The channel gain between UAV $u$ and user $k$ can be written as \cite{Azari20_transwir}

\begin{equation}\label{channel_access}
L_{u, k}^z(t) = \delta^z_u + \eta^z_u \log_{10} d_{u,k}(t) + \chi^z_u \  \ [\mathrm{dB}],
\end{equation}
where  superscript $z\in\{\mathrm{LoS}, \mathrm{NLoS}\}$ denotes a LoS or non-LoS component. 
Parameters $\delta^z_u$ and $\eta^z_u$ represent  the  reference path  loss  and  the  path  loss  exponent,  respectively.
Here, $\chi^z_b$ denotes  a zero-mean Gaussian random variable with a standard deviation $\sigma_{b,\mathrm {SF}}^z$ in dB. 

We assume that the HAPSs transmit over the orthogonal channels and also  there is no interference between the HAPSs and the UAVs (spectrum overlay access). However, multiple UAVs 
can transmit over the same channel and cause co-channel interference. Let $\omega_{\mathrm{H}}$ and $\omega_{\mathrm{U}}$ denote the total bandwidth for the HAPSs and  the UAVs, respectively. 
The total bandwidth  $\omega_{\mathrm{U}}$ (or $\omega_{\mathrm{M}}$) is divided into $|\mathcal Q_{\mathrm{U}}|$ (or $|\mathcal Q_{\mathrm{M}}|$) orthogonal channels with bandwidth $\omega_{\mathrm{U}}/|\mathcal Q_{\mathrm{U}}|$ (or $\omega_{\mathrm{M}}/|\mathcal Q_{\mathrm{M}}|$), where $\mathcal Q_{\mathrm{U}}$  (or $\mathcal Q_{\mathrm{M}}$) is the set of available channels for the UAVs (or the HAPSs).
Let $p_u$ and $g_{u, k}(t)$ denote the transmit power of UAV $u$ and the channel gain between UAV $u$ and user $k$ at time instant $t$, respectively. Therefore, the maximum achievable data rate to user $k$ provided by UAV $u$ can be expressed as 

 \begin{equation}\label{asso1}
\begin{aligned}
C_{u,k}(t)= \frac{\omega_{\mathrm{U}}}{|\mathcal Q_{\mathrm{U}}|}\log_2(1+\gamma_{u,k}(t))
 \ \ \mathrm{[bps]},
\end{aligned}
\end{equation}
where $\gamma_{u,k}(t)$ denotes the   signal to interference
plus noise ratio (SINR)   at the receiver of user $k$ associated to UAV $u$, which can be written as 

 \begin{equation}\label{asso1}
\begin{aligned}
\gamma_{u,k} (t) = \frac{{I}_{u,k}(t) p_u g_{u, k}(t)}{\sum_{u^\prime \in \mathcal{U} \backslash {u}}  p_{u^\prime} g_{u^\prime,k}(t)\rho_{u^\prime}(t) \mathbbm{1}_{(q_u(t) = q_{u^\prime}(t))} + \sigma_0^{2}} , 
\end{aligned}
\end{equation}
where $q_u(t)$ and $\sigma_0^2$ is the transmit channel of UAV $u$ at time $t$ and the   noise power, respectively. Here, $\rho_{u}(t)$ represents the load of UAV $u$ at time $t$, and the binary element 
 $I_{u,k}(t) \in \{0, 1\}$ indicates the association between UAV $u$ and user $k$ at time $t$ which can be defined as follows:

\begin{equation}\label{bi_ele_uav}
    I_{u,k}(t) = 
    \begin{cases}
    1, & \mbox{if user $k$ is associated to UAV $u$ at time $t$,}  \\
    0, & \mbox{o.w.}
    \end{cases}
\end{equation}

The achievable rate for user $k$ associated to  HAPS $m$ is given by

 \begin{equation}\label{asso1}
\begin{aligned}
C_{m,k}(t)= \frac{\omega_{\mathrm{M}}}{|\mathcal Q_{\mathrm{M}}|}\log_2(1+\gamma_{m,k}(t))
 \ \ \mathrm{[bps]},
\end{aligned}
\end{equation}
where $\gamma_{m,k}(t)$ denotes the   SINR  at the receiver of user $k$ associated to HAPS $m$, which can be defined  as 
 \begin{equation}\label{asso1}
\begin{aligned}
\gamma_{m,k} (t) = \frac{{I}_{m,k}(t) p_m g_{m, k}(t)}{\sigma_0^2} , 
\end{aligned}
\end{equation}
where $p_m$ and $g_{m,k}(t)$ denote the transmit power of HAPS $m$ and the channel gain between HAPS $m$ and user $k$, respectively.  $I_{m,k}(t) \in \{0, 1\}$ represents the association between HAPS $m$ and user $k$ at time $t$ which can be defined as follows:

\begin{equation} \label{bi_ele_hap}
    I_{m,k}(t) = 
    \begin{cases}
    1, & \mbox{if user $k$ is associated to HAPS $m$ at time $t$,}  \\
    0, & \mbox{o.w.}
    \end{cases}
\end{equation}

Let  $\mathcal K_m(t)$ and $ \mathcal K_u(t)$ denote the set of associated users to HAPS $m\in\mathcal M$ and UAV $u\in \mathcal{U}$, respectively. According to $I_{u,k}(t)$ and $I_{m,k}(t)$ defined in  \eqref{bi_ele_uav} and \eqref{bi_ele_hap}, we  can define $\mathcal K_u(t)$ and $\mathcal K_m(t) $  as follows:

\begin{equation}
    \mathcal K_u(t) = \{k | k\in\mathcal K, I_{u,k}(t) =1\},
\end{equation}

and 

\begin{equation}
    \mathcal K_m(t) = \{k | k\in\mathcal K, I_{m,k}(t) =1\}.
\end{equation}

\subsection{Load and User-ABS Association Policy}
 Now, we define the load of ABS $b\in\mathcal B$ at time instant $t$ as follows \cite{Sumudu16_TWC}:

\begin{equation}\label{load_coupled}
\begin{aligned}
&\rho_b (t) = \sum_{k\in\mathcal K_b(t)} \frac{\vartheta_k}{\zeta_k  ~ C_{b,k}(t)} 
\triangleq  f_b(\boldsymbol \rho(t)),
\end{aligned}
\end{equation}
where  $\vartheta_k$  and $\zeta_k$ are  the packet arrival rate and the mean packet size of user $k$, respectively. {{Here, ${\vartheta_k}/{\zeta_k}$  represents  the  user rate  requirement.}} Under this definition, we can consider heterogeneous users, which have different  user rate  requirements.
Vector $\boldsymbol \rho(t)=\big(\rho_1(t), \dots, \rho_{|\mathcal B|}(t)\big)$ denotes the load vector which comprises the load of all the ABSs in the system. Let  $\boldsymbol f(\boldsymbol\rho(t)) = \Big(f_1(\boldsymbol\rho(t)), \dots, f_{|\mathcal B|}\boldsymbol\rho(t)\Big)^T$. Thus, we can express \eqref{load_coupled} in the form of a vector as follows \cite{atefeh_access17}:

 \begin{equation}\label{load_coupled_vector}
\begin{aligned}
\boldsymbol \rho(t) = \boldsymbol f(\boldsymbol\rho(t)).
\end{aligned}
\end{equation}
 It is worth noting  that $f(\boldsymbol\rho(t)) $ is a standard interference function. Therefore, the non-linear load coupling equation \eqref{load_coupled_vector} can be solved by the fixed point iteration algorithm starting from an arbitrary initial   ABS load vector $\boldsymbol \rho^0>0$  as follows \cite{atefeh_icc2021}:

 \begin{equation}\label{load_coupled_min}
\boldsymbol \rho^\iota= \min \left(\boldsymbol f(\boldsymbol \rho^{\iota-1}),1\right),
\end{equation}
where  $\boldsymbol \rho^{\iota}$ denotes the  load  vector at  iteration $\iota \in \{1, \dots, N_{\mathrm{FP}}\}$, and $N_{\mathrm{FP}}$ is the total number of fixed point iterations. To ensure the system is stable, we need to guarantee loads of the ABSs not exceed the value one. 
However, in the case that a load of an ABS  $b$ exceeds the threshold one, it would drop some of its associated users to achieve $\rho_b\leq 1$ \cite{load_coupled_16_TSP}.

\textbf{Definition 1.} \textit{A function $f(\boldsymbol n)$ is called a standard interference
function, if for all $n\geq0$ the following properties are satisfied} ~\cite{SIF1995}:\textit{
\begin{enumerate}
\item Positivity: $f(\boldsymbol n) > 0,$
\item Monotonicity: $\boldsymbol n \geq \boldsymbol n^\prime\Rightarrow f(\boldsymbol n)\geq f(\boldsymbol n^\prime),$
\item Scalability: $\alpha f(\boldsymbol n) > f(\alpha \boldsymbol n)$   for $\alpha>1$,
\end{enumerate}
}

 Lemma 1 indicates that the BS load vector  $\boldsymbol \rho^{N_{\mathrm{FP}}}$ converges to the fixed point solution of \eqref{load_coupled_vector}.

\textit{Lemma 1:}  If the fixed point of \eqref{load_coupled_vector} {exists}, then it is unique, and can be iteratively obtained by  \eqref{load_coupled_min} as $N_{\mathrm{FP}}$ goes to infinity. 

\begin{proof}
According to \cite{Fehske2012}, it is proved that $ f_b(\boldsymbol \rho(t))$ is a standard interference function. Furthermore, Theorem 7 in \cite{SIF1995} prove that $\min(f_b(\boldsymbol \rho),1)$ is a standard interference function. Then, by using Theorem 2 in \cite{SIF1995}, the convergence is proved.
\end{proof}

The  assignment of the users to the ABSs needs to be addressed. Due to the mobility of the users in the system, they are expected to periodically assess their performance and make necessary adjustments. If  a user is not satisfied with its current ABS association, it may change its serving ABS and establish a new association. Therefore, new users and users that are currently experiencing an outage require to initiate new association procedures in order to be associated with new ABSs.  Given the fixed locations  and the transmit channels of the ABSs, each user is associated with an ABS based on the following user association policy:

\begin{equation}\label{ue_BS_association}
\begin{split}
b^*_k(t)  ={\mathop{\argmax }_{b\in\mathcal B}}\{p_b  g_{b,k}(t) \}.
\end{split}
\end{equation}

\subsection{Problem Formulation}

Given the described system, the objective is to maximize fairness among the users  while minimizing the load of the ABSs under the constraint of load. The optimized parameters are the UAVs' trajectories and the transmission channels.  
Here, we introduce the fairness factor to the objective function  named  Jain's fairness index, the most widely-used fairness metric in wireless networks' applications.  The  Jain's fairness index at time $t$ can be defined as follow \cite{jain1999throughput,jain1984quantitative, fairness_surv14}:

\begin{equation}\label{fairness_metric}
\mathcal F(t) = \frac{\big(\sum_{k\in\mathcal K}\bar{C}_k (t)\big)^2}{|\mathcal K| \big(\sum_{k\in\mathcal K}\bar{C}_k (t)^2\big)}.
\end{equation}
where $\bar{C}_k (t)$ is the total data rate for user $k$ until time instant $t$  be expressed as follows:

\begin{equation}\label{fairness_metric2}
\bar{C}_k (t) = \sum_{\tau\leq t} \sum_{b\in\mathcal B} C_{b,k}(\tau).
\end{equation}
The definition in \eqref{fairness_metric} reveals that the fairness index $\mathcal F(t)$ is continuous so that  a  change in a user rate results in a  change in the fairness index. Furthermore,
 it is applicable to any size of users' sets in the system. 
Besides, it is  bounded between $\frac{1}{|\mathcal K|}$ and $1$, in which a totally fair system has a Jain index of $1$ while $\frac{1}{|\mathcal K|}$ corresponds to the least fair system. 
Therefore, the higher value of the fairness index is the result of the smaller  differences among the total data rates of the users $\{\bar{C}_k (t)\}_{k\in\mathcal K}$.
Note that Jain's fairness index  takes into consideration all the users in the system, not only the users with   poor performance \cite{Gohary13}. In addition, it is mostly used for assessing long-term fairness performance. 
Furthermore,  fairness and loads of the ABSs are  unitless metrics, and they are the functions of the locations of the ABSs and the resource allocation procedure. Thus, we can combine  them to define a reward function. 
Furthermore,  the configuration of the  system can be determined by  the transmit channels of the ABSs $\boldsymbol q(t) = (q_1(t), \dots, q_{|\mathcal B|}(t))$, the locations of the ABSs $\boldsymbol Z^{\mathrm {ABS}}(t) = (\boldsymbol z_1^{\mathrm {ABS}}(t), \dots, \boldsymbol a_{|\mathcal B|}^{\mathrm {ABS}}(t))$,
and the association indicators $\boldsymbol I(t) = \{I_{u,b}\}_{b\in\mathcal B, k \in \mathcal K}$.

Our goal is to maximize  an objective function which captures both fairness and load of the ABSs.
In this regard, the optimization problem can be expressed as follows:

\begin{subequations} \label{max_prob}
\begin{align}
\max _{\substack{\boldsymbol q(t), \boldsymbol Z^{\mathrm {ABS}}(t)}}
~~ & \sum_{t\in \mathcal N}\sum_{b\in \mathcal B} \sum_{k\in\mathcal K_b(t)} \Big(\phi_{b} \mathcal F(t) + \psi_{b} (1-\rho_b(t))\Big)\\
\text { s.t. } & x_u(t)\in[x_{\mathrm{min}}, x_{\mathrm{max}}], \quad \forall u \in \mathcal U, \label{eqe1}\\
 & y_u(t)\in[y_{\mathrm{min}}, y_{\mathrm{max}}], \quad \forall u \in \mathcal U, \label{eqe2}\\
& h_u(t)\in[h_{\mathrm{min}},h_{\mathrm{max}}], \quad \forall u \in \mathcal U, \label{eqe3}\\
& q_u(t) \in \mathcal Q_{\mathrm U},   ~~\forall u \in \mathcal U, \label{chnn_cnst}\\
& \rho_b(t) = f_b(\boldsymbol\rho),~ ~ \forall b \in \mathcal B, \label{load_vct}\\
& 0\leq \rho_b(t)\leq 1, \quad \forall b \in \mathcal B,\label{load_indivi} \\
& I_{b, k}(t) \in \{0, 1\}, ~~ \forall b \in\mathcal B,  \forall k \in \mathcal K, \label{acc1}\\
& \sum_{b\in\mathcal B} I_{b,k}{(t)}\leq 1, ~~ \forall k\in\mathcal K, \label{acc2}
\end{align}
\end{subequations}
where $\phi_b$ and $\psi_b$ indicate the   weight  parameters  for the fairness index and the load of  ABS $b$ on the objective function, respectively. 
$x_{\mathrm{min}}$ and $x_{\mathrm{max}}$ are the minimum and the maximum point of horizontal ordinate in a
Cartesian coordinates of the system, respectively. 
$y_{\mathrm{min}}$ and $y_{\mathrm{max}}$ denote the minimum and the maximum point of vertical ordinate in a
Cartesian coordinates of the system, respectively. 
$h_{\mathrm {min}}$ and $h_{\mathrm {max}}$ indicate the  minimum and the maximum altitude of the UAVs, respectively.
The constraints in \eqref{eqe1}-\eqref{eqe3} determine the feasible area in the $3$D space for the locations of the UAVs at each time instant $t$ in the system.
The constraint in \eqref{chnn_cnst} represents the constraint on the set of available channels  for the UAVs.
The constraints in  \eqref{load_vct}-\eqref{load_indivi} guarantee the limitation on the load of the ABSs. 
The constraints  in \eqref{acc1}-\eqref{acc2} ensure each user $k$ is associated with at most one ABS at each time instant $t$. 

The following remarks characterize the difficulties in solving the problem formulated in \eqref{max_prob}. First, due to the presence of binary association indicators $\boldsymbol I(t) = \{I_{u,b}\}_{b\in\mathcal B, k \in \mathcal K}$ and non-convex optimization problem, the problem in \eqref{max_prob} is NP-hard. Moreover, due to the mobility of the users and the inherent highly dynamic nature of the system, the problem is very difficult to solve and it is intractable to find a globally optimal solution. Given the non-convexity and high complexity of the problem in \eqref{max_prob}, our pragmatic target is to find a high-performance solution in a reasonable amount of time.

Due to the inherent hyper-heterogeneity characteristics of SAGINs, we can use a hybrid method combination of a centralized and distributed approach. In this regard, we take advantage of both approaches. The advanced hardware processing units  with  fast computation speed and compatibility with various algorithms make to utilize the DQN algorithm in a distributed manner at the levels of the UAVs. The benefits of distributed approaches in wireless networks, such as reducing the signaling overhead and robustness to failures and attacks,  have been widely recognized in the literature \cite{atefeh_tvt17, Tang_JSAC22}. 
For the centralized part, we assume that there is a  cloud radio access network (C-RAN) for sharing information regarding the data rates of the users to calculate fairness \cite{tsipi2022unsupervised}. In this regard, at the beginning of each time slot, the C-RAN broadcasts the calculated Jain's fairness index to the UAVs. Then, the UAVs are allowed to employ the broadcasted  data   and process their own information. Thus, Once a new UAV is launched  into the system,  it will first listen to the beacons, and then will start the action selection process.
At the end of the time slot, each ABS calculates the data rates of its associated users and send these values to the C-RAN. 
This procedure results in a more adaptive and flexible system and can reap the benefit of both centralized and distributed approaches.

\section{Deep Reinforcement Learning-Based  Link Optimization}

In this section, we first present an overview of Q-learning  
and DQN. Then,  a DQN-based scheme for resource management and trajectory design is proposed. It utilizes  both load and fairness using a replay memory method to achieve the formulated objective function which is described in \eqref{max_prob}. 
Since the load balancing and fairness optimization problem is a high dimensional and high state/action problem, we must employ novel and state-of-the-art methods such as  
DQN algorithm. 
The proposed algorithm enables UAVs to learn the entire
network environment to adjust their positions jointly with determining their transmit channel.
Finally, we present a detailed state, action, and reward function design.

\subsection{Learning Model}
The use of learning methods in wireless networks has received unprecedented attention, in which they show significant  improvements over traditional mechanisms.  Among them, reinforcement learning (RL),e.g., Q-learning, has achieved remarkable success for different problems in complex and highly dynamic systems.  
In RL, agents interact with the environment and take action. Then, they observe the consequences of their actions which can lead to learning their optimal policies.  
This success  is due to the procedure of effectively finding the optimal policy for  a finite Markov decision process (MDP). 
The MDP can be expressed as a four-tuple $<\mathcal S, \mathcal A, \mathcal R, P>$, where $\mathcal S$ implies the observable environment states, $\mathcal A$ is the set of alternative actions. $\mathcal R$ indicates the reward function for taking action $a \in \mathcal A$ in state $s \in \mathcal S$ \cite{fu_Collection_IoT21, adachi_20_tvt, Danhao22_cletter}. $P: \mathcal S \times \mathcal A \times R \rightarrow [0, 1]$ is the state transition probability distribution function. The actions of an agent are selected based on a policy $\pi: \mathcal S \rightarrow \mathcal A$, which is a mapping from the state space to the
action space. RL algorithms aim at learning an optimal policy $a = \pi(s) \in \mathcal A$. 
The agent will adjust its policy $\pi$ in order to maximize its long-term expected return $E[G_n]$, which is given by \cite{Flow_level19}:

\begin{equation}
G(n) \delequal\sum_{k = 0}^{\infty}  \gamma^k 
 R({n+k}),
\end{equation}
where $G(n)$ is the accumulated discounted reward, and  $0\leq\gamma\leq1$ is  the  discount factor of future reward, which  makes trade-off immediate
rewards with the rewards generated in future time instants.  
Let $Q_{\pi}(s, a)$ represent 
 the action-value function of executing action $a$ under state $s$ following policy $\pi$ as the average cumulative discount reward. The action-value function  can be defined as follows:

\begin{equation}
Q_{\pi}(s, a) = E_{\pi} [ G(n)| S(n) = s, A(n) = a].
\end{equation}

In Q-learning, an agent in a state  takes an action  and observes a reward. To select an action, it has two options: choosing an action with the highest Q-value or selecting a random action. Then, it updates the Q-table based on the observed reward. 
Q-learning is an off-policy reinforcement learning algorithm, in which it gradually improves its strategies with its accumulation of experience and  strives to find the best action at any state. 
To evaluate the quality of an action-state pair, the algorithm  updates the Q-value function using the  Bellman equation according to  the weighted average of the current  Q-value function and the reward as follows \cite{Shahriar_JCN20}:

\begin{equation}
\begin{split}
\begin{aligned}
        Q\big(s(t), a(t)\big) \leftarrow &Q\big(s(t), a(t)\big) + \alpha \Big(r(t) + \\
        & \gamma \max_{a}{Q\big(s({t + 1}), a\big)} - Q\big(s(t), a(t)\big)\Big)
\end{aligned}
\end{split}
\end{equation}
where $Q(s(t), a(t))$ is the Q-value function for state $s(t)$ and action $a(t)$ at time $t$, and $\alpha$  is the learning rate. Note that this table-based reinforcement learning method is suitable for problems  with limited action-state space. Despite the great empirical success of  Q-learning, it is less applicable to real-world problems.  This is due to the fact that most   real-world problems  are complex and have large or continuous  action-state spaces so that they remain unaddressed hindering the deployment of  Q-learning-based solutions.  Therefore, using the table-based Q-learning algorithms to solve those problems is challenging and it is not feasible to apply them directly to complex and highly dynamic environments. 
To practically use RL algorithms for problems with large or continuous action-state space, the function approximation method can be employed. 
DQN is an extension of the Q-learning algorithm that combines deep neural networks with a reinforcement learning framework. 
In the DQN algorithm, a deep neural network is employed to approximate Q-values instead of using a Q-table to represent $Q\big(s(t), a(t)\big)$ which can allow us to deal with large action-state spaces. However, 
the Q-Network will  take the state as an input and return the expected Q-values for every action. 
Thus, $Q\big(s(t), a(t); \theta\big)$ is the estimated
Q-value function during the iterative process, which is
approximated by the neural network with the weights of $\theta$.
In the training process, $Q\big(s(t), a(t); \theta\big)$ is updated by adjusting weights $\theta$.
In our system, we choose to represent the state as a multi-dimensional array that contains the information of the $3$D location and transmit channel of a UAV, in which the location is normalized and a one-hot decoder is used for the channel. The action space 
includes the movement direction 
and the transmit channel of  UAVs.  
We will discuss further the elements of our proposed model in Section \ref{solution_DQN_sec}.

\begin{figure*}[t]
\centering\resizebox{6in}{3.8in} {\includegraphics
{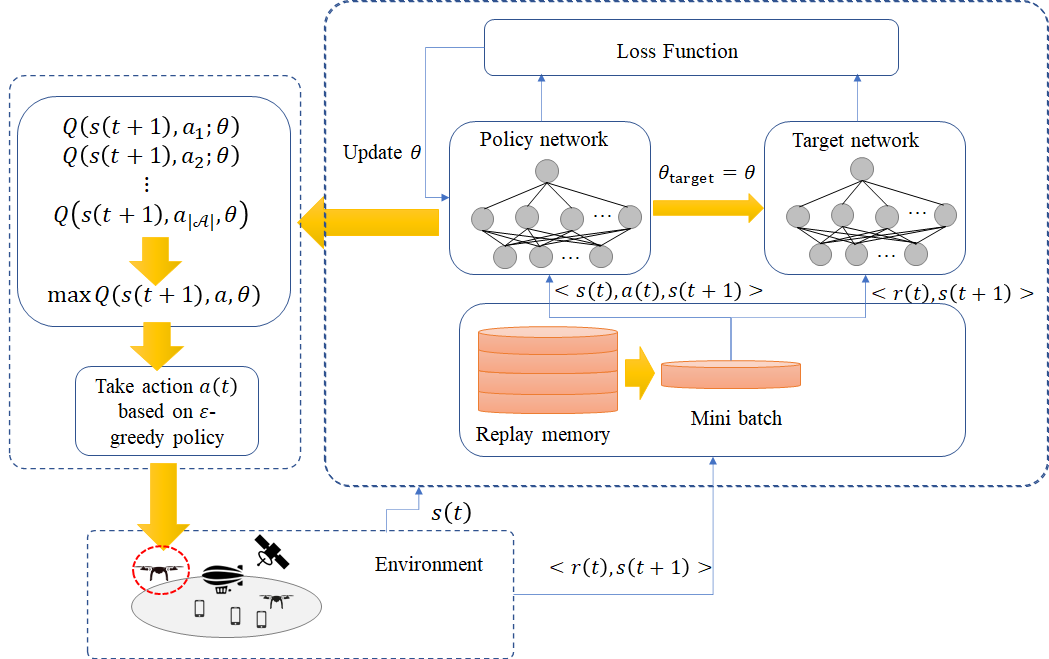}}
\caption{Overview of the DQN structure.} \label{dqn_str}
\end{figure*}

Given the environment, each UAV learns to take the best action depending on the current state during the training phase. 
In the Q-learning method, it updates its Q-table according to the  returned reward value which shows how good it is to take a given action in a given state. 
On the other hand, in a DQN method, the model is not represented using a table, but it is represented by a set of weights and biases in a neural network referred to as a Q-network compared to the Q-table. The DQN is composed of two neural networks including the policy and the target network.  To train the models, the weights and biases of policy and target networks are initialized randomly.
To optimize the learning process, a replay memory, {shown in Fig. \ref{dqn_str}}, is incorporated for updating the Q-network \cite{Yitong_iot22}. 
Replay memory is an efficient technique to reuse  previous experiences, and it  allows the agent  to learn from earlier memories. In this regard, experiences are
stored in a memory buffer with a fixed size. When the replay memory 
is full, the oldest memories are erased \cite{liu2018effects}. Furthermore, to 
update the agent's parameters, a random batch of experiences is sampled from the replay memory.
Using replay memory can address the issues relevant to  the temporal correlations and enhances data usage and computation efficiency. 
%
%
It stores the agent's instances which include the past state, selected action, reward, and the next state given the selected action.
Let $< s(t), a(t), r(t), s({t+1})>$ represent a sample from the replay memory. 
Then, the agent randomly samples a batch from the replay memory.   To take an action, an $\epsilon$-greedy model is used, which allows the agent to explore its action space, and it can be defined as follows:

\begin{equation}\label{eps_action_select}
   a(t)\!=\!\begin{cases}
 \mbox{a random action}, & \mbox{with probability } \epsilon\\
\argmax_{a\in\mathcal A}{Q(\!s(t+1),\!a;\!\theta)}, & \mbox{with probability } 1-\epsilon,
  \end{cases}
\end{equation}
where $\epsilon>0$ is an exploring ratio which  is adaptively updated according to   
the following expression:

\begin{equation} 
\epsilon \leftarrow \epsilon_{\mathrm{end}} + (\epsilon_{\mathrm{start}} - \epsilon_{\mathrm{end}})         \exp(-\tau / \epsilon_{\mathrm{decay}}),
\end{equation}
where $\epsilon_{\mathrm{start}}$ and $ \epsilon_{\mathrm{end}}$ denote the  start value and the end value for the $\epsilon$-greedy threshold, respectively. $\epsilon_{\mathrm{decay}}$ is the threshold decay, and $\tau$ indicates as many as steps done for selecting an action based on the $Q(s(t), a; \theta)$. 
The output of the policy network $Q(s(t), a; \theta)$ is used as the decision of the agent, whereas the output of the target network is used to update the networks through computing a loss function which compares the outputs of the policy and target networks. To choose action $a(t)$,  at time instant $t$, state $s(t)$ is fed into the neural network with weights $\theta$, and $a(t)$ is obtained as $a(t) = \argmax_{a} Q(s(t), a; \theta)$ or through a random selection according to \eqref{eps_action_select}, where $ Q(s(t), a; \theta)$ denotes the outputs of the neural network corresponding to all possible
actions $a$. After taking action $a(t)$, the agent received reward $r(t)$ and moves to the next state $s({t+1})$. Then, the DQN is trained by minimizing the prediction error of $Q(s(t), a(t); \theta)$ using the loss function $L_{\delta}(y, \hat{y})$. 
We use the Huber loss to minimize the loss so that when the loss is small, it acts as the mean squared error (i.e., L2 loss), whereas the loss is large, it acts as the mean absolute error (i.e., L1 loss) which makes it more robust to outliers for the noisy estimations of the neural networks \cite{sun2020adaptive}. The loss is calculated over a batch of transitions sampled from the replay memory as follows \cite{Tang_JSAC22}:

\begin{equation}\label{loss_func}
L_{\delta}(y, \hat{y}) = \begin{cases}
 \frac{1}{2} (y - \hat{y})^2, & \mbox{if } |y - \hat{y}| \leq \delta \\
\delta |y-\hat{y}| - \frac{1}{2} \delta^2, & \mbox{otherwise},
  \end{cases}
\end{equation}
where $y$ and $\hat{y}$ denote the output of the learning system, i.e.,  $Q(s(t), a(t); \theta)$, and the target value, respectively. The target value $\hat{y}$ can be estimated as

\begin{equation}\label{target_value}
\hat{y} = r(t) + \gamma \max_{a} Q(s({t+1}), a(t); \theta_{\mathrm{target}}).
\end{equation}
Parameter $\delta > 0$ specifies the threshold at which to change between delta-scaled L1 and L2 loss. Here, the target value $\hat{y}$ is computed based on the obtained reward and predicted discounted reward $\gamma \max_{a} Q(s({t+1}), a(t); \theta_{\mathrm{target}})$ given by the target network, where $\theta_{\mathrm{target}}$ denotes the weights of the target network.  
Unlike the policy network, which continuously updates its weights based on the observed rewards and actions, the weights of the target network are not updated iteratively. Instead, they are periodically updated by copying the weights of the policy network after a specified time interval \cite{Tiankui_safe_dqn21}.
Note that to calculate the loss function,  the agent picks  a random batch from the replay memory rather than using a single sample which leads to improving the learning stability. 
After calculating the loss, it is fed into an optimizer to update the weights and biases of the neural networks. 
In our model, we use RMSprop optimizer which is an adaptive algorithm to evaluate gradient updates \cite{graves2013generating}. The update rules in RMSprop are as follows:

\begin{equation} \label{rmsprop_avg}
E[g^2]_t = \eta E[g^2]_{t-1} + (1-\eta) g^2_t,
\end{equation}

\begin{equation}\label{update_theta}
\theta({t+1}) = \theta(t) - \frac{\alpha}{\sqrt{e[g^2]_t + \epsilon}} g_t,
\end{equation}
where $g_t$  is the gradient at time $t$. Parameters $\eta$ and $\alpha$ denote the constant forgetting factor and the initial learning rate, respectively. $\theta$ is the weights and biases in the neural networks, respectively. Then, these updates are applied to the model.  Fig. \ref{dqn_str} illustrates a structure of a DQN approach for a UAV in an aerial network.

\subsection{DQN-Assisted UAV Operation Algorithm} \label{solution_DQN_sec}

\begin{figure}[t]
\hspace*{-0.5cm}
\centering\resizebox{3.7in}{2.5in} {\includegraphics
{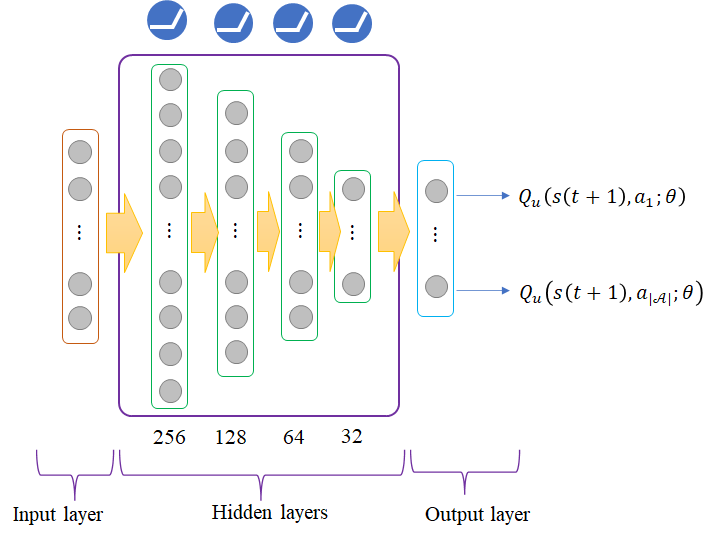}}
\caption{Structure of the policy network.} \label{dqn_layers}
\end{figure}

In the proposed approach, UAVs are seen as agents which interact with the system environment in a sequence of discrete time instances. At each time $t$, each UAV $u$ observes the state $s_{u}(t)$, takes action $a_{u}(t)$ and receives  the reward $r_{u}(t)$. Then, it moves to the new state $s_{u}(t+1)$ at time $t+1$. 
Furthermore, each UAV utilizes a replay memory $\mathcal D_u$ with a certain   capacity to store the transition sample  $<s_u(t), a_u(t), r_u(t), s_u({t+1})>$.
In the context of the described problem, we define the state $s_{u}(t)$, action $a_{u}(t)$, and reward $r_{u}(t)$ for UAV $u$ at time instant $t$ as follows:

\begin{itemize}
    \item State representation $\boldsymbol s_{u} (t)$: each UAV $u\in\mathcal U$ determines  state  $\boldsymbol  s_{u} (t)$ from  its location,  i.e., $\boldsymbol z_u^{\mathrm {ABS}}(t) = (x_u(t), y_u(t), h_u(t))$, and transmit channel $q_u(t)$.  Here, we introduce an encoder to encode the UAV's transmitted channel information into a unique vector using a one-hot code.
    In one-hot encoding, a variable is represented by  a one-hot vector, e.g., $1 \rightarrow [0, 1, 0, 0], 4 \rightarrow [0, 0, 0, 0, 1]$. 
     More precisely, one-hot encoding is a process, which is used to convert  categorical variables into a suitable form feeding to Q-networks \cite{goodfellow2016deep}.  Thus, the UAV translates each state into a $0-1$ string, and then it sends the state vector  into the Q-network. The UAV's location is normalized by the minimum and the maximum values of the UAV's altitude, point of horizontal, and vertical ordinate. 
    In this regard, the state of UAV $u$ can be expressed as  $\boldsymbol  s_{u} (t) = \{\Bar{\boldsymbol x}_u(t), \Bar{q}_u(t)\}$. Here, $\Bar{\boldsymbol x}_u(t) = (\frac{x_u(t)}{x_{\mathrm{max}}- x_{\mathrm{min}}},  \frac{y_u(t)}{y_{\mathrm{max}}- y_{\mathrm{min}}}),  \frac{h_u(t)}{h_{\mathrm{max}}- h_{\mathrm{min}}})$ denotes the normalized value of the UAV's location $\boldsymbol z_u^{\mathrm {ABS}}(t)$, and $\Bar{q}_u(t)$ is the one-hot encoded of the transmit channel $q_u(t)\in\mathcal Q$.


    
    \item Action: For each UAV $u\in\mathcal U$, action $\boldsymbol  a_u(t) = \{z_u(t), q_u(t)\}$, where $z_u(t) \in \mathcal{Z}$ and $q_u(t) \in \mathcal Q$  denote the movement direction and transmit channel of UAV $u$ at time $t$, respectively. The set of movement directions is defined as
    \begin{equation}
          \mathcal{Z} = \{ \mathrm{up,down,left,right,forward,backward,}\\
     \mathrm{fixed} \}.
        \end{equation}
Therefore, the action space for each UAV $u\in \mathcal U$ can be described as 
    \begin{equation}
\mathcal A\!=\!\{\boldsymbol  a_u(t) | \boldsymbol  a_u(t) = \{z_u(t), q_u(t)\},z_u(t) \in \mathcal{Z}, q_u(t) \in \mathcal Q\}.
    \end{equation}
    \item Reward: the reward is the objective of the dynamic resource management and trajectory design problem. This function is consistent with
the mathematical formulation of our optimization problem. Thus,  for each UAV $u\in \mathcal U$, the reward function is related to load and fairness, and according to  {\eqref{max_prob}}, it can be defined as follows:
\begin{equation}\label{reward}
r_u(t) = \phi_{u} \mathcal F(t) + \psi_{u} (1-\rho_u(t)).
\end{equation}
After taking action $a_u(t)$ by UAV $u$, it receives the reward $r_{u}(t)$ and  moves to the new state $s_u(t+1)$.
\end{itemize}
To approximate the Q-function values, each UAV utilizes two deep networks for policy and target networks with the same four fully connected layers while they have different weights and biases.  From {Fig. \ref{dqn_layers}}, we can see that the  neural network is composed of three parts, including the input layer, hidden layers, and output layer. 
In our model, we employ $4$ hidden layers with $256$, $128$, $64$, and $32$ nodes. For the activation function, ReLU is selected.   Furthermore, after each layer except the output layer,  we apply layer normalization and drop out  with   probability $0.2$. The layer normalization technique can enhance  the performance and stability of neural networks \cite{batch_norm_2015}. It  normalizes the inputs to a layer, thereby enabling the utilization of higher learning rates and faster convergence.
The dropout technique operates by randomly disconnecting the connections between neurons in connected layers 
 based on a certain dropout rate to reduce the dependency between neurons \cite{srivastava2014dropout}. 
The input of the neural network corresponds to the state of the UAV, and the output corresponds to action-value approximations. For the policy network, the input is the current state-action pair $(s_u(t), a_u(t))$  and the output is the predicted value $Q(s_u(t), a_u(t); \theta)$. For the target network, the input is the next state $s_u(t+1)$ and the output is the maximum Q-value of the next state-action pair so that the target value of $(\boldsymbol s_u(t), \boldsymbol a_u(t))$ for UAV $u$ can be calculated  as follows:

\begin{equation}\label{target_value}
\begin{split}
\begin{aligned}
   \hat{y}_u(\boldsymbol s_u(t), \boldsymbol a_u(t))& =\\
&r_u(t) + \gamma \max_{a} Q_u(\boldsymbol s_u({t+1}), \boldsymbol a_u(t); \theta_{\mathrm{target}}).
\end{aligned}
    \end{split}
\end{equation}
It is important to note that each UAV has its own DQN network, with its own unique set of neural network weights, distinct from the other UAVs.   Algorithm \ref{alg_dqn} presents the  pseudocode for our proposed   approach.  It is noteworthy that we apply the heuristic-based initialization for our proposed DQN-based approach. In this algorithm, the horizontal location of a new UAV is determined based on the furthest distances from the other BSs in the system \cite{halim_ini} (lines \ref{halim_alg_l1}-\ref{halim_alg_end}). 
\begin{algorithm}[tb!]
\caption{:   DQN-based algorithm for   $3$D trajectories and resource managmenet in aerial HetNets}
\label{alg_dqn}
\begin{algorithmic}[1]
\STATE \textbf{Input}:  a differentiable Q-value function parameterization $Q_u(s, a; \theta)$ and $\theta_{\mathrm{target}} = \theta$ , $\forall u\in\mathcal U$ 

\STATE\textbf{Initialization}: a replay memory $\mathcal D_u, \forall u \in\mathcal U$

\textit{Initialization of the UAVs' locations}: 

\STATE $\boldsymbol f^{\mathrm{ABS}}(0) \Leftarrow \boldsymbol f_{\mathrm H}$,   $\mathcal {B}^* \Leftarrow \mathcal H$, $u = 0$, $\boldsymbol f^{\mathrm {UAV}}_u(0) = \{\}, \forall u \in \{1, \dots, |\mathcal U|\}$  \label{halim_alg_l1}

\WHILE{$u <|\mathcal U|$}
\FOR{$\forall l \in \mathcal L$}
\FOR{$\forall b \in \mathcal B^*$} \label{Halim_alg_e1}
\STATE $r_{l,b} = ||\boldsymbol f_{l} - \boldsymbol f^{\mathrm ABS}_b||$
\ENDFOR \label{Halim_alg_e2}
\STATE $r_{l}^{\mathrm{min}} = \min \limits_{b\in\mathcal B^*}~{r_{l,b}}$   \label{Halim_alg_e3}
\ENDFOR
\STATE $l^* = \argmax \limits_{l\in \mathcal L} r_{l}^{\mathrm {min}}$ \label{Halim_alg_e4}
\STATE $\mathcal L \Leftarrow \mathcal L \backslash \{l^*\}$,  $\boldsymbol f^{\mathrm{ABS}}(0)\Leftarrow \boldsymbol f^{\mathrm{ABS}}(0)
\cup  \{\boldsymbol f_{l^*}\}$, $\mathcal B^* \Leftarrow \mathcal B^*\cup \{u\}$, $\boldsymbol f^{\mathrm {UAV}}_u(0) = \boldsymbol f_{l^*}$ \label{Halim_alg_e5}
\STATE $u \Leftarrow u + 1$
\ENDWHILE \label{halim_alg_end}
\vspace{0.2cm}

\textit{Learning procedure:}
\FOR{episode$:=1, N_{\mathrm{episode}}$}
\WHILE{$t<N$}
\STATE $t \leftarrow t+1$
\FOR{\textbf{each} $k \in \mathcal K$}
\STATE Update $\boldsymbol z_k^{\mathrm {UE}}(t)$ based on the random walk mobility model described in Section  \ref{ue_mob_sec}
\STATE Associate user $k$ to an ABS according to \eqref{ue_BS_association}
\STATE Update the user association indicators according to  
\eqref{bi_ele_uav} and \eqref{bi_ele_hap}  
\ENDFOR

\FOR{\textbf{each} $u \in \mathcal U$}
\STATE Select an action according to \eqref{eps_action_select}
\STATE Update the location $\boldsymbol{z}_u^{\mathrm{ABS}}(t)$ based on  $a_u(t)$ and \eqref{eq_UP_LO}
\STATE Calculate reward $r_u(t)$ according to \eqref{reward} and move to the next state $s_u(t + 1)$
\STATE Store the transition sample \\ $<s_u(t), a_u(t), r_u(t), s_u(t+1)>$ into $\mathcal D_u$
\STATE Sample a stochastic minibatch of samples from $\mathcal D_u$
\STATE Compute target value according to \eqref{target_value}
\STATE Update weights $\theta$ by minimizing the loss \eqref{loss_func}
\STATE Update the target network parameters  $\theta_{\mathrm{target}}$ every $N_{\mathrm{T}}$ steps as $\theta_{\mathrm{target}} = \theta$ 
\ENDFOR
\ENDWHILE
\ENDFOR
\end{algorithmic}
\end{algorithm}%
%
We define the set of all predefined locations for the UAVs as $\mathcal L$ and a single location in this set as $l$. 
The two-dimensional ($2$D)  coordinate of a location $l$ is represented by $\boldsymbol{f}_l$  while the vector  composed of the locations of the ABSs in the system is represented by  $\boldsymbol{f}^{\mathrm{ABS}}(0)$. 
The initial ABS locations, $\boldsymbol{f}^{\mathrm{ABS}}(0)$, are determined by the HAPSs, as $\boldsymbol{f}^{\mathrm{ABS}}(0)  \Leftarrow  \boldsymbol{f}_{\mathrm {H}}$, where $\boldsymbol f_{\mathrm H} = (\boldsymbol f_1^{\mathrm {ABS}}, \dots, \boldsymbol f_{|\mathcal H|}^{\mathrm {ABS}})$ represents the $2$D locations of all HAPSs. $f^{\mathrm {UAV}}_u(0)$ denotes the  selected  location for UAV $u$. 
The set of current ABSs in the system,  $\mathcal B^*$, is initialized with the set of HAPSs.
At each iteration, the algorithm determines the initial location of a new UAV. The 2D distance, $r_{l,b}$, between each ABS $b$ in $\mathcal B^*$ and location $l$ is calculated (lines \ref{Halim_alg_e1}-\ref{Halim_alg_e2}).  Then,   the distance between location $l$ and the nearest ABS in set $\mathcal B^*$, which is denoted by $r_l^{\mathrm{min}}$ is calculated  (line \ref{Halim_alg_e3}).
Finally, the location $l^*$ with the farthest distance from ABSs in  $\mathcal B^*$ is  selected as the UAV location, denoted as $l^*$ (line \ref{Halim_alg_e4}). Then, the location $l^*$ is removed from $\mathcal L$ and its coordinate  $\boldsymbol{f}_{l^*}$  is added to the ABS locations in $\boldsymbol{f}^{\mathrm {ABS}}(0)$ (line \ref{Halim_alg_e5}).
To initialize the transmit channels of the UAVs, we adopt a random selection, in which the UAVs choose their channels from a uniform distribution, i.e. $\pi_{u, q} = \frac{1}{|Q|}$ for $\forall u \in \mathcal{U}$ and $\forall q \in \mathcal Q$, where $\pi_{u, q}$ is the 
probability assigned channel $q \in \mathcal Q$ for UAV $u\in\mathcal U$.

\section{Simulation Results}

\begin{table}[tb!] 
\vspace{0.2cm}
\begin{center}
\caption{System-Level Simulation Parameters } \label{t_sim_par}
\begin{tabular}{|p{3in}|p{1.in}|p{0.8in}|} \hline
\multicolumn{3}{|c|}{\textbf{System Parameters}} \\ \hline \hline
\multicolumn{2}{|p{1.8in}|}{\textbf{Parameter}}& \textbf{Value}  \\  \hline
\multicolumn{2}{|p{1.8in}|}{Height of the HAPS}  & $20$ km \cite{wrc19-al}\\  \hline
\multicolumn{2}{|p{1.8in}|}{$h_{\mathrm{min}}, h_{\mathrm{max}}$} & {$22.5$ m, $150$ m} \\ \hline
\multicolumn{2}{|p{1.8in}|}{Height of users}  & $1.5$ m\\  \hline


\multicolumn{2}{|p{1.8in}|}{Carrier frequency UAV and HAPS} & $28$, $2.11$ {GHz} \\  \hline

\multicolumn{2}{|p{1.8in}|}{$|\mathcal Q_{\mathrm{U}}|, |\mathcal Q_{\mathrm{M}}|$} & $4, 1$ \\  \hline
\multicolumn{2}{|p{1.8in}|}{$\omega_{\mathrm{U}}, \omega_{\mathrm{M}}$} & $56, 14$ MHz \\  \hline
\multicolumn{2}{|p{1.8in}|}{Noise power spectral density} & $-174$ dBm/Hz \\  \hline
\multicolumn{2}{|p{1.8in}|}{Number of HAPSs} & $1$ \\  \hline
\multicolumn{2}{|p{1.8in}|}{Height of HAPSs } & $20$ km \\ \hline
\multicolumn{2}{|p{1.8in}|}{Total number of iterations
($N$)} & $5740$\\  \hline
\multicolumn{2}{|p{1.8in}|}{$T_s$} &  $1$ sec\\  \hline
\multicolumn{2}{|p{1.8in}|}{Fixed point iterations ( $N_{\mathrm{FP}}$)} & $500$\\  \hline
\multicolumn{2}{|p{1.8in}|}{$\rho^0_b$} & $0.5$\\  \hline
\multicolumn{2}{|p{1.8in}|}{$\alpha,\beta, \xi$} & $0.1, 750, 8$\\  \hline
\multicolumn{2}{|p{1.8in}|}{$v^{\mathrm{min}}_\mathrm{UE}, v^{\mathrm{max}}_\mathrm{UE}$}& $0$, $1.3$ m/sec  \\ \hline
 \multicolumn{2}{|p{1.8in}|}{$v_{\mathrm{U}}$}
& {$10$ m/sec} \\ \hline
\multicolumn{2}{|p{1.8in}|}{${\vartheta_k}/{\zeta_k} $} & $1.8$ Mbps \\ \hline
\multicolumn{2}{|p{1.8in}|}{$\phi_{b}, \psi_{b}$} & $0.5, 0.5$ \\ \hline
\multicolumn{2}{|p{1.8in}|}{Transmit power of the HAPS and UAVs} & $43$, $24$ dBm \\ \hline
 \multicolumn{2}{|p{1.8in}|}{Reference path loss} & $61.4$ \cite{hamed_ahmadi2020} \\ \hline
  \multicolumn{2}{|p{1.8in}|}{Path loss exponent  LoS/NLoS} & $2$,  $3$ \\ \hline
  \multicolumn{2}{|p{1.8in}|}{Shadowing standard deviation  LoS/NLoS} &  $5.8$,  $8.7$ \\ \hline
  \multicolumn{2}{|p{1.8in}|}{$\gamma$} & $0.999$\\ \hline
    \multicolumn{2}{|p{1.8in}|}{Batch size} & $128$\\ \hline
    \multicolumn{2}{|p{1.8in}|}{Repaly memory size} & $5000$\\ \hline
    \multicolumn{2}{|p{1.8in}|}{Minimum repaly memory size} & $264$\\ \hline
        \multicolumn{2}{|p{1.8in}|}{Target network frequency update ($N_{\mathrm{T}}$)} & $10$\\ \hline
    \multicolumn{2}{|p{1.8in}|}{$\epsilon_{\mathrm{start}}, \epsilon_{\mathrm{end}}, \epsilon_{\mathrm{decay}}$} & $0.9, 0.5, 200$\\ \hline

 \end{tabular}
\end{center}
\vspace*{-0.5cm}
\end{table}

In the simulation scenario, a  $1000 \times 1000~\mathrm{m}^2$ area is considered, and a set of users are uniformly distributed throughout the area.
Furthermore, a HAPS is located at the center of the area at a height of $20$ km from the ground \cite{wrc19-al}. 
Table \ref{t_sim_par} summarizes the system parameters employed in the simulations.  The simulation results are obtained by averaging over numerous independent runs with variations using practical configurations. 
Furthermore, the performance of our proposed DQN scheme is evaluated through comparison with several benchmark algorithms
as follows:

\begin{itemize}
    \item \textit{DQN-No HAPS}: To demonstrate the advantages of incorporating HAPS,  the DQN-No HAPS scheme is implemented. In this approach, only UAVs are employed for data transmission to the users, without the presence of HAPS. The UAVs optimize their trajectories and transmit channels using the proposed DQN algorithms. 
    \item \textit{Q-learning}: In the Q-learning approach, both UAVs and HAPSs are deployed to provide service to the users. The $2$D positions  and the transmit channels of the UAVs are optimized using a Q-learning technique. The altitude of the UAVs is set at $h_{\mathrm{max}}$. 
   \item  \textit{Q-learning-No HAPS}: In this benchmark algorithm, no HAPSs are employed, and only UAVs provide service for users. The UAVs optimize their trajectories and transmit channels using a Q-learning technique, flying at the  fixed altitude of $h_{\mathrm{max}}$.
\end{itemize}

\begin{figure}[tb!]
\centering\resizebox{3.7in}{2.2in}  {\includegraphics
{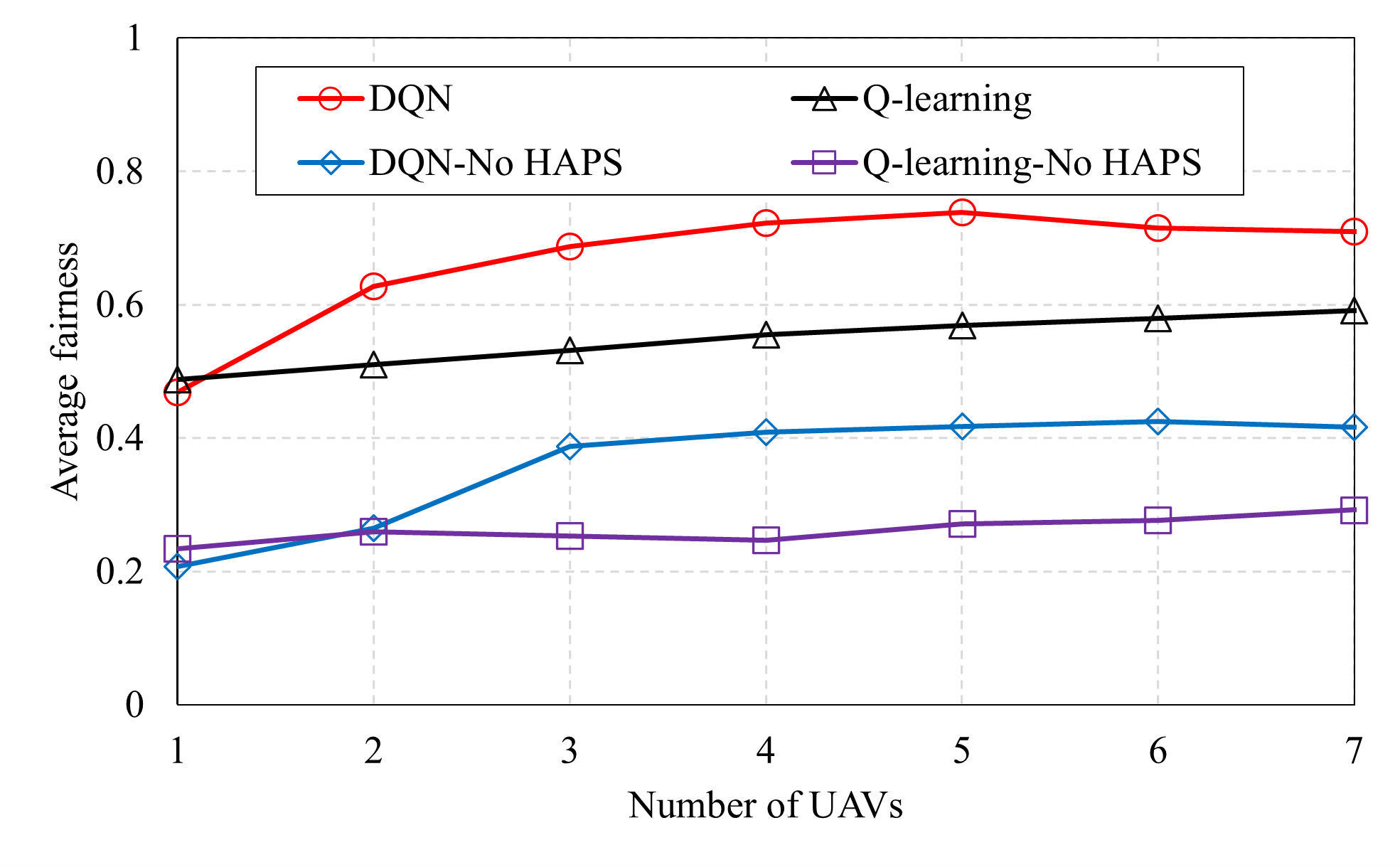}
}
\caption{{Average fairness versus the number of UAVs for a system with $200$ users.}} \label{fair_uav}
\end{figure}

Fig. \ref{fair_uav} presents the impact of the number of UAVs on Jain's fairness index defined in \eqref{fairness_metric}  which is used to quantify the distribution of resources among the users in the system. It shows that as the number of UAVs increases, Jain's fairness index improves. The main reason is that the additional UAVs provide more resources and coverage to the network which leads to a more fair distribution of resources among users and ensures an enhanced user experience and improved network performance. Furthermore, the DQN approach significantly outperforms the benchmark algorithms. This improved performance is due to the ability of the DQN approach to learn from experience and adapt to changing conditions in the system. However, for a system with a single UAV, the Q-learning (or Q-learning-No HAPS) approach slightly performs better than DQN (or DQN-No HAP) mechanism. This is due to the fact that in the Q-learning algorithm, the altitude of the UAV is set at the maximum altitude. Thus, it can cover more area and support more users due to providing a high probability of LoS links. 

\begin{figure}[tb!]
\centering\resizebox{3.7in}{2.2in}  {\includegraphics
{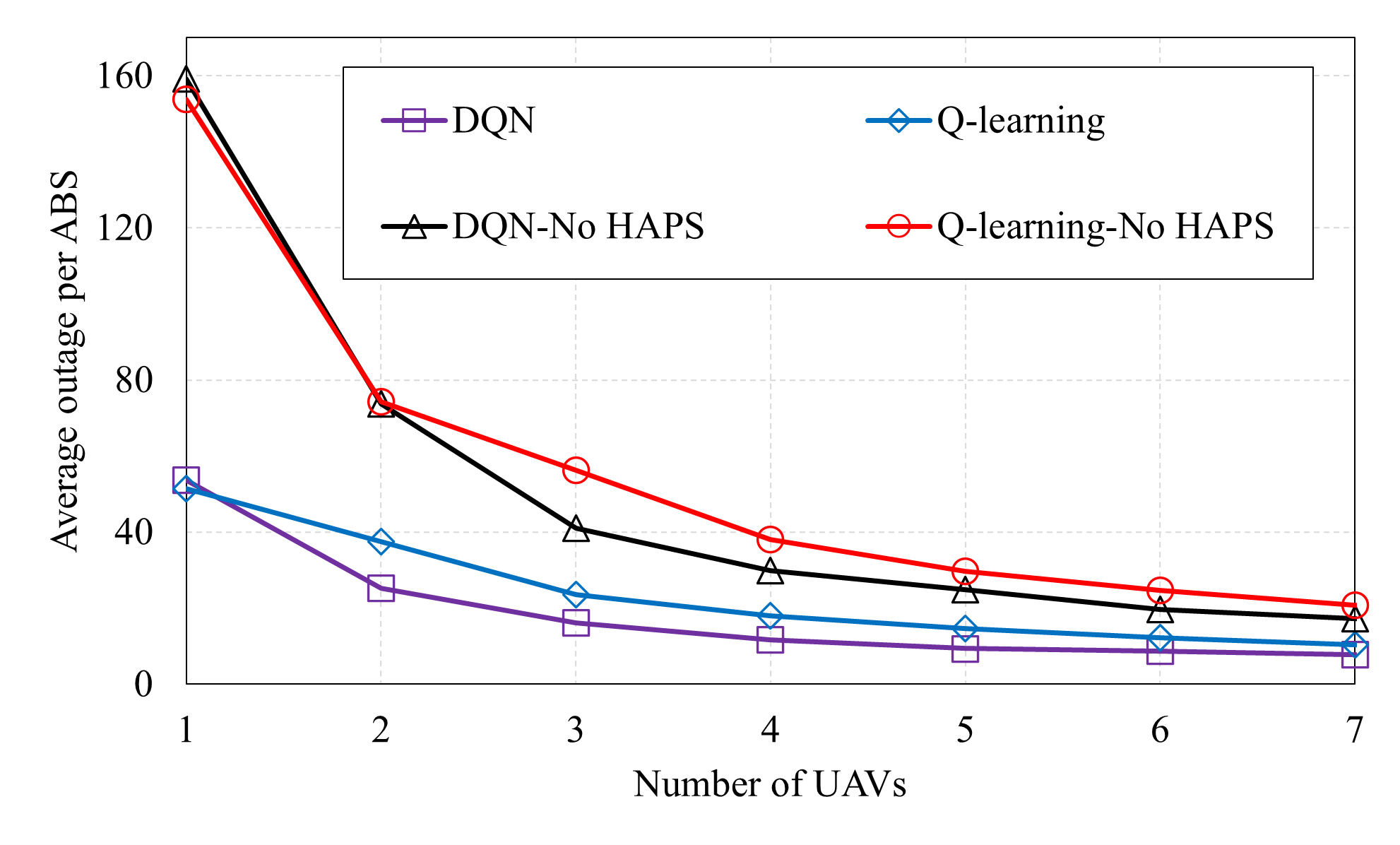}
}
\caption{{Average  outage per ABS  versus the number of UAVs for a system with $200$ users.}} \label{drop-uav}
\end{figure}

As shown in Fig. \ref{drop-uav}, the performance of the proposed DQN approach is compared with the benchmark algorithms in terms of outage users. The figure  illustrates the relationship between the number of UAVs and the average number of outage users and the scalability of our proposed DQN approach. 
Outage users refer to users  which experience disconnection or a drop in the received data rate. 
 Therefore, it is  imperative for network operators and service providers to effectively monitor and manage the number of outage users to ensure the sustainability and reliability of the network. In addition, the number of outage users is a critical performance metric and can be used to assess the efficacy of network optimization strategies and resource allocation algorithms. From Fig. \ref{drop-uav}, it can be observed that, as the number of UAVs increases, the average number of outage users per ABS decreases for all methods. However, the proposed DQN approach outperforms the benchmark algorithms, demonstrating its effectiveness  in reducing the number of outage users and improving service coverage.
 This result highlights the effectiveness of the proposed DQN approach in improving the resource allocation and $3$D trajectory design for UAV-based communication systems.
 The reason for the decrease in the average number of outage users as the number of UAVs increases is due to the improved resource allocation and more efficient utilization of the available UAVs. With a larger number of UAVs, the loads are balanced over the ABSs, and thus  more users can be served which reduces the number of users without service. Additionally, having more UAVs with effective interference management  methods enables a more flexible design and better serves the users. Note that increasing the number of ABSs in the system may cause more interference if the resource is not allocated properly. 
  The benchmark algorithms without HAPSs which only employ UAVs, degrade the performance in terms of outage users. This is due to the limited coverage area of UAVs which leads to inadequate service quality for some users, especially in dense areas.  In contrast, the proposed DQN approach which leverages both UAVs and HAPSs, provides  a larger coverage area and improved service quality, thus it reduces the number of outage users.

\begin{figure}[tb!]
\centering\resizebox{3.7in}{2.2in}  {\includegraphics
{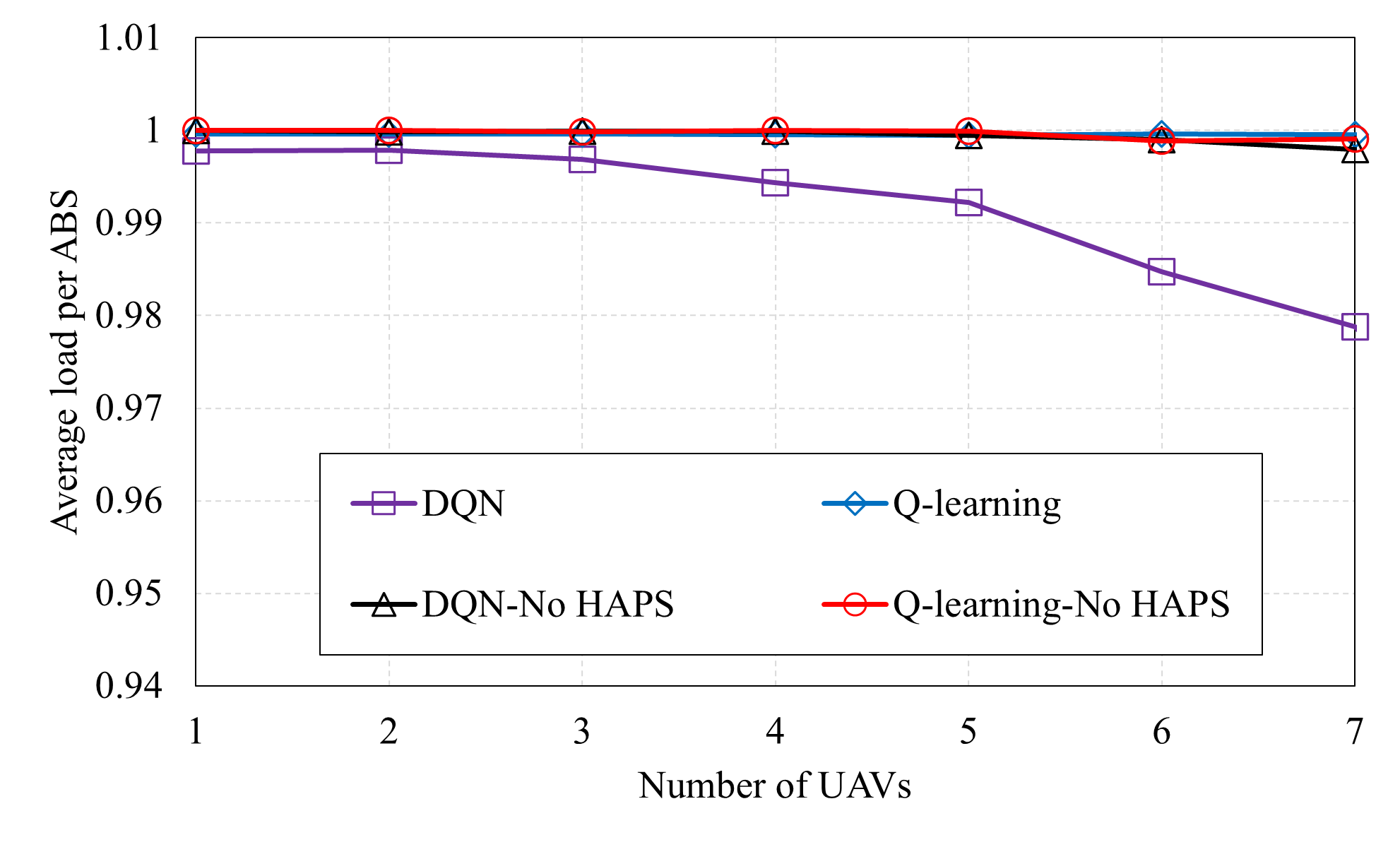}
}
\caption{{Average load per ABS  versus the number of UAVs for a system with $200$ users.}} \label{load_uav}
\end{figure}

Fig. \ref{load_uav} shows the average load per ABS as the number of UAVs increases. The results indicate that in the proposed approach, as the number of UAVs increases, the average load per ABS decreases.
This behavior  helps to alleviate the overloading of the  ABSs and ensures efficient and stable service provided to the users.
Furthermore, the results indicate that the benchmark methods are not capable of effectively balancing the load in the system, in which with the increasing number of UAVs, there is a limited decrease in average load.
 Specifically, for the dense deployment of UAVs, the proposed DQN approach shows an improvement in terms of load balancing compared to the benchmark algorithms. The gap between the proposed approach and the benchmark algorithms  becomes larger as the number of UAVs increases which demonstrates the effectiveness of the DQN approach in 
 ensuring a balanced distribution of load among the ABSs
 in densely deployed UAVs scenarios by managing the resource and optimizing the $3$D  locations of the UAVs.

\begin{figure}[tb!]
\centering\resizebox{3.7in}{2.2in}  {\includegraphics
{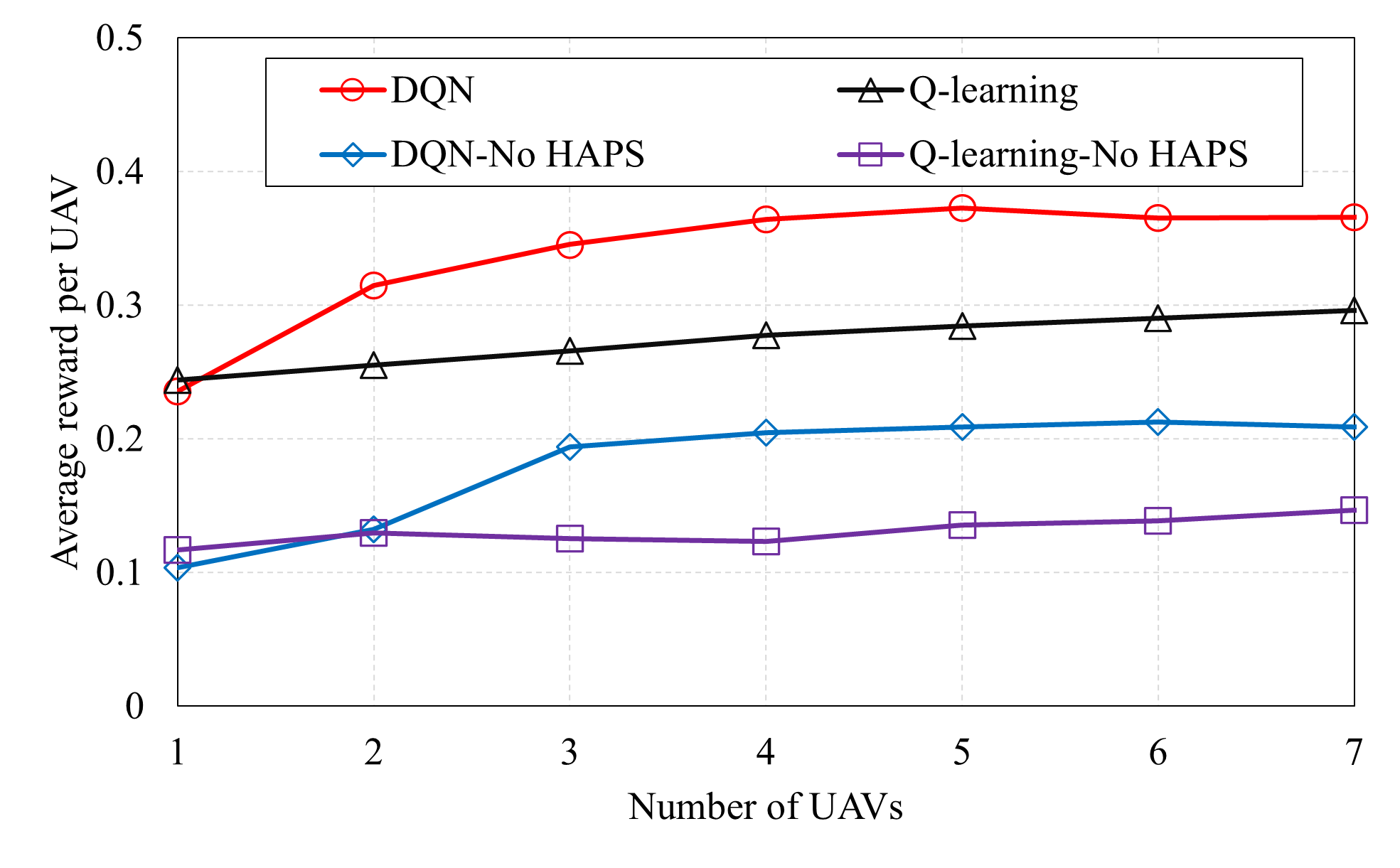}
}
\caption{{Average reward per UAV  versus the number of UAVs for a system with $200$ users.}} \label{reward_uav}
\end{figure}

The average reward, defined in \eqref{reward}, per UAV as a function of the number of UAVs is depicted in Fig. \ref{reward_uav}.   The average reward can be considered  as a suitable performance metric to assess the success of the methods in optimizing the system's objective.   A higher reward value indicates that the algorithm is successful in satisfying the  objective function, while a lower reward value indicates that the algorithm is encountering difficulties to achieve  desired outcomes.
As illustrated in Fig.  \ref{reward_uav}, the  DQN approach  outperforms the benchmark algorithms in terms of the average reward achieved by  the UAVs. 
The improvement in reward achieved by the proposed DQN approach is a result of the decreased load of the ABSs and improved fairness among users. By optimizing the UAV's trajectories and transmission channels, the DQN approach ensures an equitable distribution of resources, thereby improving both load balancing and fairness. Since the reward function captures both load and fairness, thus improving both parameters  results in a higher overall reward compared to the benchmark algorithms.
However, it should be noted that for scenarios involving a single UAV, the Q-learning approach performs slightly better than the DQN approach. This is due to the fact that in the Q-learning method, the altitude of the UAV is set at the maximum altitude, enabling it to cover a larger area with a high probability of LoS.
Additionally, in the scenario with a single UAV, due to the lack of interference, setting the altitude of the UAV at the maximum altitude results in improving the performance of the Q-learning method. {However, employing more UAVs may increase interference in the system, which requires critical factors such as load balancing and fairness provisioning to be optimized dynamically and intelligently.} 
Additionally, the scenarios without the utilization of the HAPS result in a decreased reward compared to the scenarios that employ the HAPS. The main reason is that the HAPS provides an additional layer of support for service coverage, which leads to improved fairness.

\begin{figure}[tb!]
\centering\resizebox{3.7in}{2.2in}  {\includegraphics
{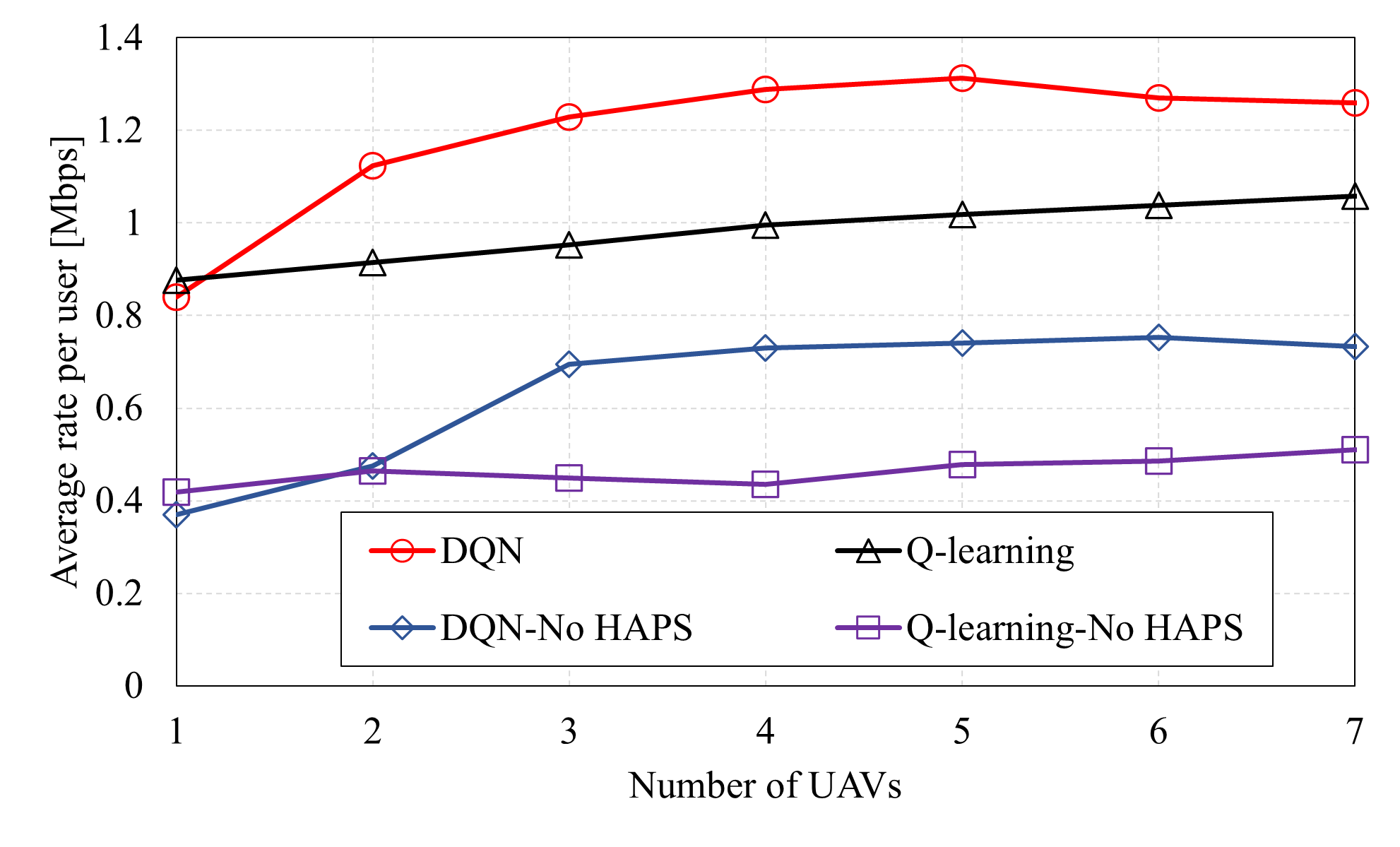}
}
\caption{{Average rate per user  versus the number of UAVs for a system with $200$ users.}} \label{rate_uav}
\end{figure}

Fig. \ref{rate_uav} illustrates the average rate per user versus the number of UAVs deployed in the system.  This figure shows a comparison of the performance of the proposed DQN approach with the benchmark algorithms and provides insights into the impact of the number of UAVs on the system performance in terms of users' rates. It can be observed that with increasing the number of UAVs, the average rate per user tends to improve. This is due to providing more resource for the users, and thus they have more opportunities to select their serving ABSs which lead to offloading outage users from highly loaded ABSs to lightly loaded ABSs. 
In addition, the DQN approach manages  interference efficiently in the system and this can lead to an increase in the number of users served by the ABSs which results in higher user rates. Similarly, by optimizing the location of UAVs and resource allocation, it is possible to reduce interference and improve user rates.
Furthermore, this figure shows the integration of the HAPS and UAVs can improve the user rate significantly compared to conventional aerial communication systems. For instance, the proposed DQN approach enhances the user rate up to about  $77$\%  compared to the DQN-No HAPS method for $2$ UAVs. It is important to note that the improvement in user rate depends on the deployment scenario, resource allocation,  and the number of ABSs used.

\begin{figure}[tb!]
\centering\resizebox{3.7in}{2.2in}  {\includegraphics
{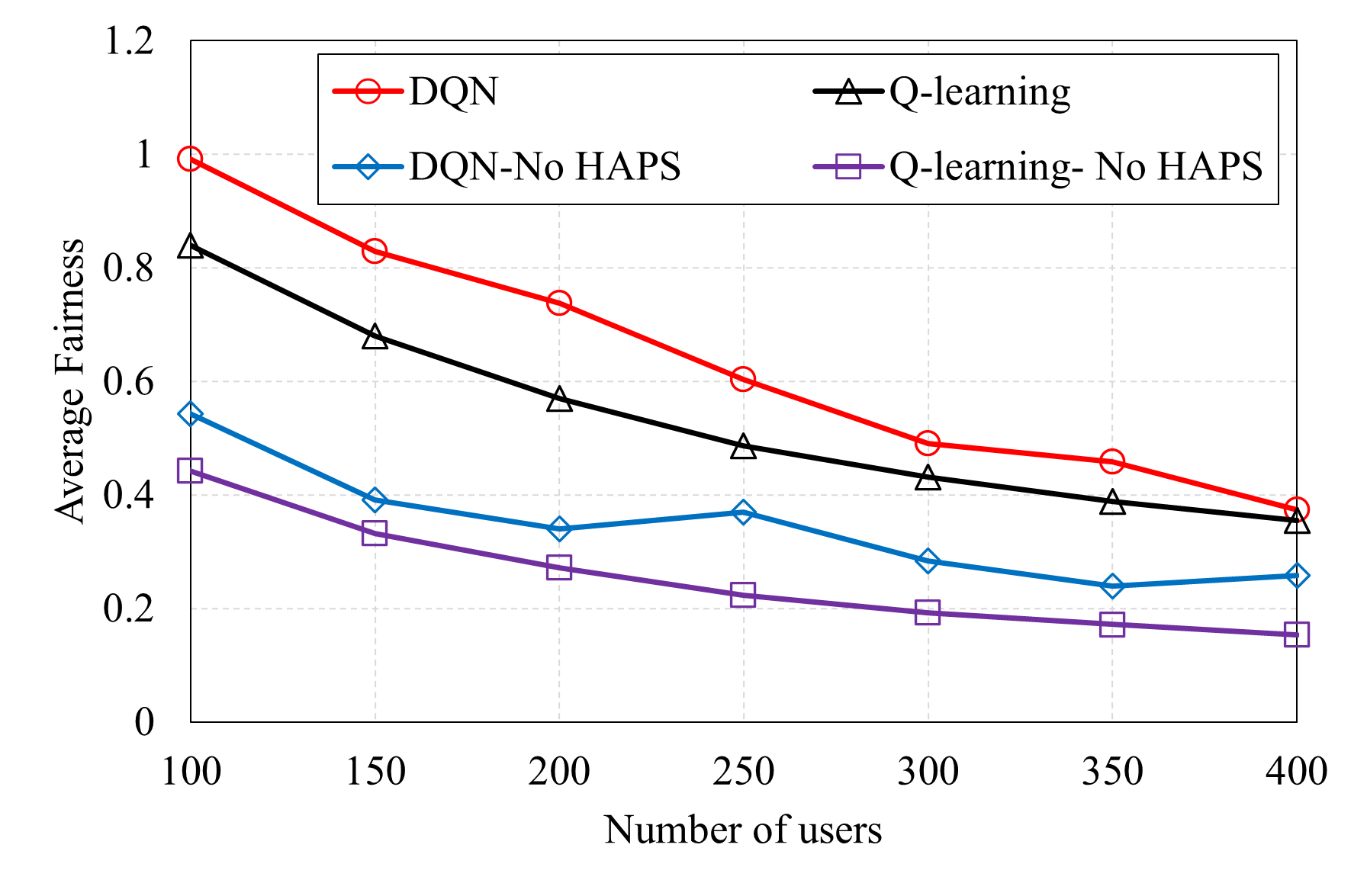}
}
\caption{{Average fairness versus the number of users for a system with $5$ UAVs.}} \label{fair_ue}
\end{figure}

Fig. \ref{fair_ue} shows the performance of the DQN approach and the benchmark algorithms in terms of Jain's fairness index versus different numbers of users. This figure can measure the distribution of resource among the users in the system. We can observe that the DQN scheme achieves improved performance in terms of fairness compared to the benchmark algorithms. This  can provide valuable insight into the scalability, flexibility, and ability of the DQN method to allocate  resources fairly  for a varying number of users  based on the states of the system.
Furthermore, the performances of all methods decrease as the number of users in the system increases. This is due to the fact that as the number of users increases, the availability of resources in the system  becomes limited. Thus, it shows the importance of effectively and fairly allocating resources among the users. 
Furthermore, in the  absence of the HAPS,  only the UAVs provide service for the users which can lead to a decrease in Jain's fairness index as resources are not distributed fairly among users. 

\begin{figure}[tb!]
\centering\resizebox{3.7in}{2.2in}  {\includegraphics
{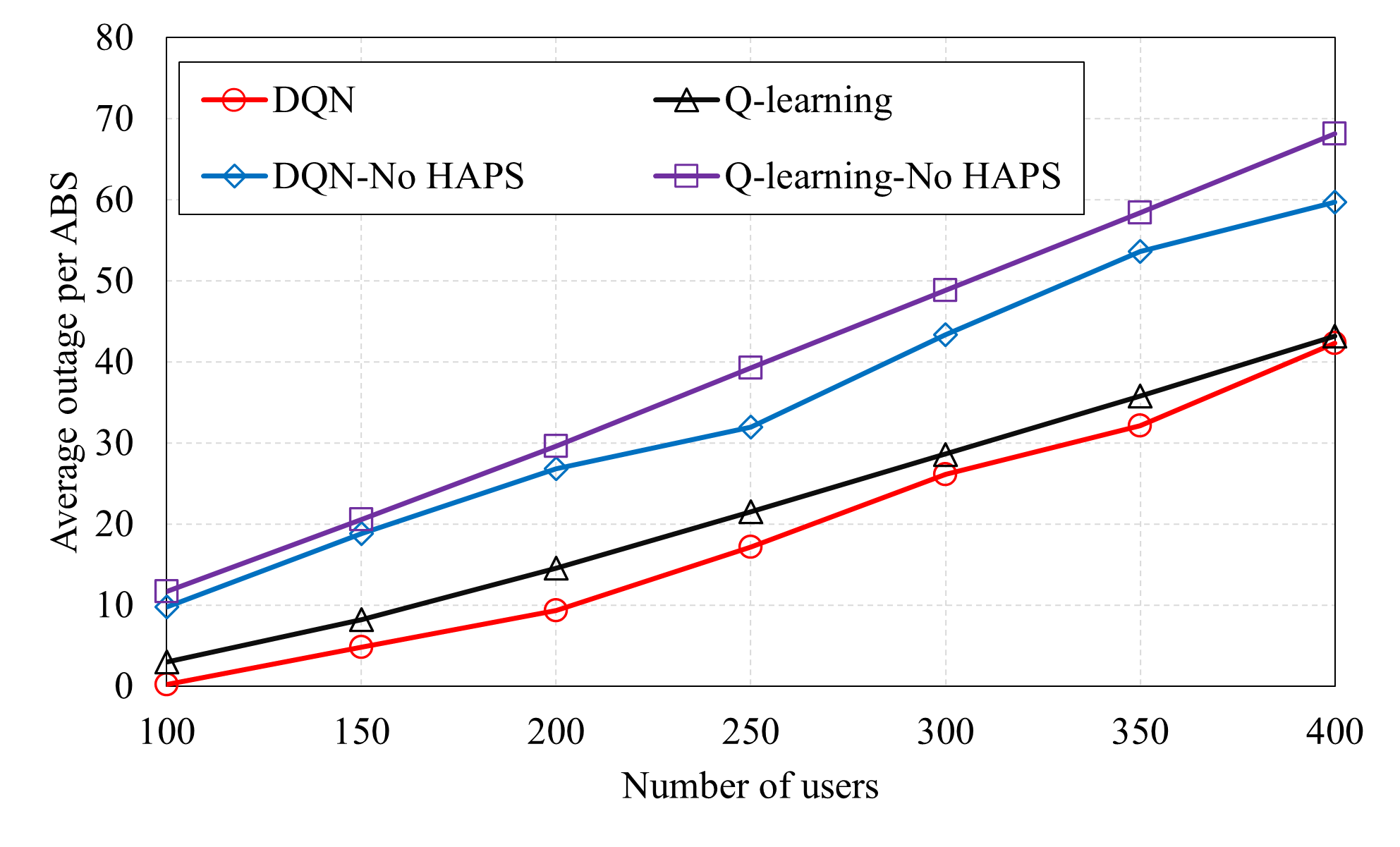}
}
\caption{{Average outage per ABS versus the number of users for a system with $5$ UAVs.}} \label{drop_ue}
\end{figure}

Fig. \ref{drop_ue}, illustrates the average number of outage users for the DQN method and the benchmark algorithms. It can be seen that the DQN approach yields better performance  compared to the benchmark algorithms. Obviously, for a fixed number of ABSs, as the number of users in the system increases, the demand for resources also increases, potentially leading to  a higher number of outage users.
Moreover,  Fig. \ref{drop_ue} demonstrates the  contribution of the HAPS  to the reduction of outage users. The deployment of HAPSs  has the potential to significantly decrease the number of outage users and can help to alleviate resource scarcity in the dense system. In the absence of HAPSs, the system relies solely on UAVs, which can lead to limited network coverage, resulting in a higher number of outage users. 

\begin{figure}[tb!]
\centering\resizebox{3.7in}{2.2in}  {\includegraphics
{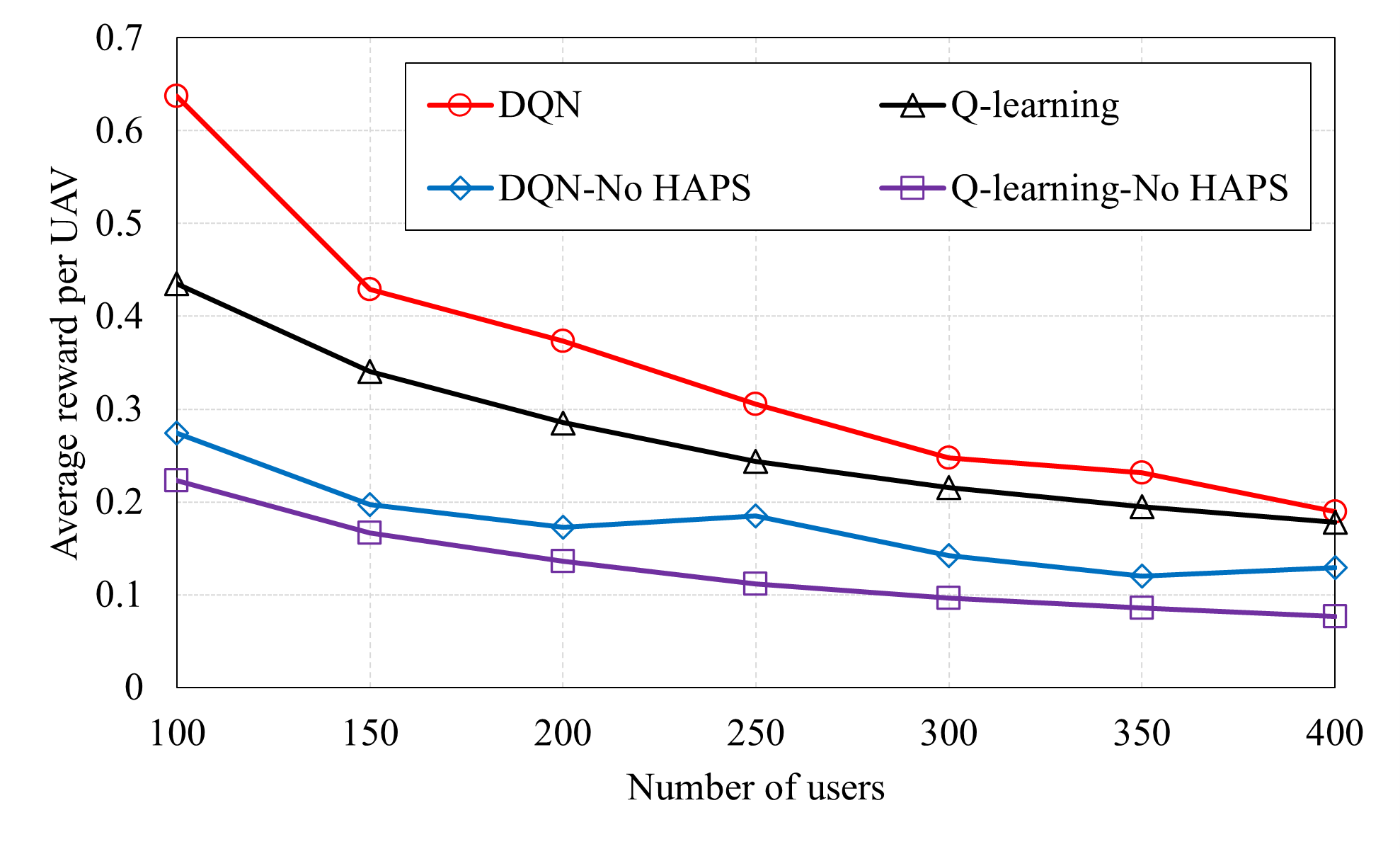}
}
\caption{{Average reward per UAV versus the number of users for a system with $5$ UAVs.}} \label{reward_ue}
\end{figure}

In Fig. \ref{reward_ue},  the average reward is plotted versus the number of users in the system to evaluate the performance of all the methods under different load conditions. As the number of users in the system increases, the UAVs are faced with a greater challenge in balancing the load and distributing resources fairly and efficiently. We can observe that the DQN algorithm demonstrates higher rewards compared to the benchmark algorithms due to its improved performance in load balancing and fairness. Compared to the benchmark algorithms, the DQN algorithm achieves better load balancing by dynamically adjusting resource allocation through channel allocation and $3$D trajectory design  based on the current state of the system.

\section{Conclusion}

In this paper, we have addressed an important problem of joint trajectory and resource management design in HAPS-UAV-enabled heterogeneous networks composed of HAPSs and UAVs as ABSs. To solve the problem, we have employed a DQN algorithm which is able to handle the complexity of the problem. Moreover,  we have utilized a fixed-pint iteration method to find the load of ABSs. 
Simulation results have shown that the integration of HAPSs and UAVs can significantly improve the performance of the network  compared to conventional communication systems and a Q-learning-based mechanism in terms of fairness, user rate, and outage.

\section{Acknowledgement}
This work was supported by the High-Throughput and
Secure Networks Challenge program of National Research Council Canada under Grant No. CH-HTSN-418.




\bibliographystyle{IEEEtran}
\bibliography{IEEEabrv,./paper.bib}

\begin{thebibliography}{10}
\providecommand{\url}[1]{#1}
\csname url@samestyle\endcsname
\providecommand{\newblock}{\relax}
\providecommand{\bibinfo}[2]{#2}
\providecommand{\BIBentrySTDinterwordspacing}{\spaceskip=0pt\relax}
\providecommand{\BIBentryALTinterwordstretchfactor}{4}
\providecommand{\BIBentryALTinterwordspacing}{\spaceskip=\fontdimen2\font plus
\BIBentryALTinterwordstretchfactor\fontdimen3\font minus
  \fontdimen4\font\relax}
\providecommand{\BIBforeignlanguage}[2]{{%
\expandafter\ifx\csname l@#1\endcsname\relax
\typeout{** WARNING: IEEEtran.bst: No hyphenation pattern has been}%
\typeout{** loaded for the language `#1'. Using the pattern for}%
\typeout{** the default language instead.}%
\else
\language=\csname l@#1\endcsname
\fi
#2}}
\providecommand{\BIBdecl}{\relax}
\BIBdecl

\bibitem{GSMA2022}
\BIBentryALTinterwordspacing
GSMA, ``{High Altitude Platform Systems - Towers in the Skies},'' Tech. Rep.
  Feb, 2022. [Online]. Available:
  \url{https://www.gsma.com/futurenetworks/wp-content/uploads/2022/02/HAPS-Towers-in-the-skies-draft-v-2.1-clean.pdf}
\BIBentrySTDinterwordspacing

\bibitem{Hao_WCSP20}
G.~Hao, W.~Ni, H.~Tian, and L.~Cao, ``Mobility-aware trajectory design for
  aerial base station using deep reinforcement learning,'' in \emph{2020
  International Conference on Wireless Communications and Signal Processing
  (WCSP)}, 2020, pp. 1131--1136.

\bibitem{Guo_IWCMC19}
J.~Guo, Y.~Huo, X.~Shi, J.~Wu, P.~Yu, L.~Feng, and W.~Li, ``{3D} aerial vehicle
  base station {(UAV-BS)} position planning based on deep {Q}-learning for
  capacity enhancement of users with different {QoS} requirements,'' in
  \emph{2019 15th International Wireless Communications \& Mobile Computing
  Conference (IWCMC)}, 2019, pp. 1508--1512.

\bibitem{Wang_JOCIN20}
L.~Wang, K.~Wang, C.~Pan, X.~Chen, and N.~Aslam, ``Deep {Q}-network based
  dynamic trajectory design for {UAV}-aided emergency communications,''
  \emph{Journal of Communications and Information Networks}, vol.~5, no.~4, pp.
  393--402, 2020.

\bibitem{Liu20_TVT}
Q.~Liu, L.~Shi, L.~Sun, J.~Li, M.~Ding, and F.~Shu, ``Path planning for
  {UAV}-mounted mobile edge computing with deep reinforcement learning,''
  \emph{IEEE Transactions on Vehicular Technology}, vol.~69, no.~5, pp.
  5723--5728, 2020.

\bibitem{atefeh_iot22}
A.~Hajijamali~Arani, M.~M. Azari, P.~Hu, Y.~Zhu, H.~Yanikomeroglu, and
  S.~Safavi-Naeini, ``Reinforcement learning for energy-efficient trajectory
  design of {UAVs},'' \emph{IEEE Internet of Things Journal}, vol.~9, no.~11,
  pp. 9060--9070, 2022.

\bibitem{Tang20_highmob}
F.~Tang, Y.~Zhou, and N.~Kato, ``Deep reinforcement learning for dynamic
  uplink/downlink resource allocation in high mobility {5G HetNet},''
  \emph{IEEE Journal on Selected Areas in Communications}, vol.~38, no.~12, pp.
  2773--2782, 2020.

\bibitem{Tang_JSAC22}
F.~Tang, H.~Hofner, N.~Kato, K.~Kaneko, Y.~Yamashita, and M.~Hangai, ``A deep
  reinforcement learning-based dynamic traffic offloading in space-air-ground
  integrated networks {(SAGIN)},'' \emph{IEEE Journal on Selected Areas in
  Communications}, vol.~40, no.~1, pp. 276--289, 2022.

\bibitem{atefeh_icc2021}
A.~H. Arani, P.~Hu, and Y.~Zhu, ``Re-envisioning space-air-ground integrated
  networks: Reinforcement learning for link optimization,'' in \emph{ICC 2021 -
  IEEE International Conference on Communications}, 2021, pp. 1--7.

\bibitem{Yuan22}
S.~Yuan, F.~Hsieh, S.~Rasool, E.~Visotsky, M.~Cudak, and A.~Ghosh,
  ``Interference analysis of {HAPS} coexistence on terrestrial mobile
  networks,'' in \emph{2022 IEEE Wireless Communications and Networking
  Conference (WCNC)}, 2022, pp. 2494--2499.

\bibitem{Jia21_JSAC}
Z.~Jia, M.~Sheng, J.~Li, D.~Zhou, and Z.~Han, ``Joint {HAP} access and {LEO}
  satellite backhaul in {6G}: Matching game-based approaches,'' \emph{IEEE
  Journal on Selected Areas in Communications}, vol.~39, no.~4, pp. 1147--1159,
  2021.

\bibitem{Ahmadinejad22}
H.~Ahmadinejad and A.~Falahati, ``Forming a two-tier heterogeneous air-network
  via combination of high and low altitude platforms,'' \emph{IEEE Transactions
  on Vehicular Technology}, vol.~71, no.~2, pp. 1989--2001, 2022.

\bibitem{APY_Access2021}
A.~H. Arani, P.~Hu, and Y.~Zhu, ``{Fairness-Aware Link Optimization for
  Space-Terrestrial Integrated Networks: A Reinforcement Learning Framework},''
  \emph{IEEE Access}, vol.~9, pp. 77\,624--77\,636, 2021.

\bibitem{APY2022}
------, ``{UAV-Assisted Space-Air-Ground Integrated Networks: A Technical
  Review of Recent Learning Algorithms},'' \emph{arXiv:2211.14931 [eess.SY]},
  2022.

\bibitem{Cao21}
X.~Cao, B.~Yang, C.~Yuen, and Z.~Han, ``{HAP}-reserved communications in
  space-air-ground integrated networks,'' \emph{IEEE Transactions on Vehicular
  Technology}, vol.~70, no.~8, pp. 8286--8291, 2021.

\bibitem{Swaminathan21}
S.~R, S.~Sharma, N.~Vishwakarma, and A.~S. Madhukumar, ``{HAPS}-based relaying
  for integrated space-air-ground networks with hybrid {FSO/RF} communication:
  A performance analysis,'' \emph{IEEE Transactions on Aerospace and Electronic
  Systems}, vol.~57, no.~3, pp. 1581--1599, 2021.

\bibitem{mobility2002survey}
T.~Camp, J.~Boleng, and V.~Davies, ``A survey of mobility models for ad hoc
  network research,'' \emph{Wireless Communications and Mobile Computing},
  vol.~2, no.~5, pp. 483--502, 2002.

\bibitem{Shibata22_acess}
Y.~Shibata, W.~Takabatake, K.~Hoshino, A.~Nagate, and T.~Ohtsuki, ``Two-step
  dynamic cell optimization algorithm for {HAPS} mobile communications,''
  \emph{IEEE Access}, vol.~10, pp. 68\,085--68\,098, 2022.

\bibitem{antenna2008}
J.~{Holis} and P.~{Pechac}, ``Elevation dependent shadowing model for mobile
  communications via high altitude platforms in built-up areas,'' \emph{IEEE
  Trans. Antennas Propag.}, vol.~56, no.~4, pp. 1078--1084, 2008.

\bibitem{hamed_ahmadi2020}
G.~{Fontanesi}, A.~{Zhu}, and H.~{Ahmadi}, ``Outage analysis for
  millimeter-wave fronthaul link of {UAV}-aided wireless networks,'' \emph{IEEE
  Access}, vol.~8, pp. 111\,693--111\,706, 2020.

\bibitem{Azari20_transwir}
M.~M. {Azari}, G.~{Geraci}, A.~{Garcia-Rodriguez}, and S.~{Pollin},
  ``{UAV}-to-{UAV} communications in cellular networks,'' \emph{IEEE Trans.
  Wireless Commun.}, vol.~19, no.~9, pp. 6130--6144, 2020.

\bibitem{Sumudu16_TWC}
S.~Samarakoon, M.~Bennis, W.~Saad, and M.~Latva-aho, ``Dynamic clustering and
  on/off strategies for wireless small cell networks,'' \emph{IEEE Transactions
  on Wireless Communications}, vol.~15, no.~3, pp. 2164--2178, 2016.

\bibitem{atefeh_access17}
A.~Hajijamali~Arani, M.~J. Omidi, A.~Mehbodniya, and F.~Adachi, ``Minimizing
  base stations' on/off switchings in self-organizing heterogeneous networks: A
  distributed satisfactory framework,'' \emph{IEEE Access}, vol.~5, pp.
  26\,267--26\,278, 2017.

\bibitem{load_coupled_16_TSP}
R.~L.~G. Cavalcante, S.~Sta\'{n}czak, J.~Zhang, and H.~Zhuang, ``Low complexity
  iterative algorithms for power estimation in ultra-dense load coupled
  networks,'' \emph{IEEE Transactions on Signal Processing}, vol.~64, no.~22,
  pp. 6058--6070, 2016.

\bibitem{SIF1995}
R.~D. Yates, ``A framework for uplink power control in cellular radio
  systems,'' \emph{IEEE J. Sel. Areas Commun.}, vol.~13, no.~7, pp. 1341--1347,
  1995.

\bibitem{Fehske2012}
A.~J. Fehske and G.~P. Fettweis, ``Aggregation of variables in load models for
  interference-coupled cellular data networks,'' in \emph{2012 IEEE Int. Conf.
  Commun. (ICC)}, 2012, pp. 5102--5107.

\bibitem{jain1999throughput}
R.~Jain, A.~Durresi, and G.~Babic, ``Throughput fairness index: An
  explanation,'' in \emph{ATM Forum contribution}, vol.~99, no.~45, 1999.

\bibitem{jain1984quantitative}
R.~K. Jain, D.-M.~W. Chiu, W.~R. Hawe \emph{et~al.}, ``A quantitative measure
  of fairness and discrimination,'' \emph{Eastern Research Laboratory, Digital
  Equipment Corporation, Hudson, MA}, 1984.

\bibitem{fairness_surv14}
H.~{SHI}, R.~V. {Prasad}, E.~{Onur}, and I.~G. M.~M. {Niemegeers}, ``Fairness
  in wireless networks:issues, measures and challenges,'' \emph{IEEE Commun.
  Surv. Tutor.}, vol.~16, no.~1, pp. 5--24, 2014.

\bibitem{Gohary13}
A.~B. Sediq, R.~H. Gohary, R.~Schoenen, and H.~Yanikomeroglu, ``Optimal
  tradeoff between sum-rate efficiency and {Jain's} fairness index in resource
  allocation,'' \emph{IEEE Transactions on Wireless Communications}, vol.~12,
  no.~7, pp. 3496--3509, 2013.

\bibitem{atefeh_tvt17}
A.~H. Arani, A.~Mehbodniya, M.~J. Omidi, F.~Adachi, W.~Saad, and
  I.~G{\"u}ven{\c{c}}, ``Distributed learning for energy-efficient resource
  management in self-organizing heterogeneous networks,'' \emph{IEEE
  Transactions on Vehicular Technology}, vol.~66, no.~10, pp. 9287--9303, 2017.

\bibitem{tsipi2022unsupervised}
L.~Tsipi, M.~Karavolos, and D.~Vouyioukas, ``An unsupervised machine learning
  approach for {UAV}-aided offloading of {5G} cellular networks,'' in
  \emph{Telecom}, vol.~3, no.~1.\hskip 1em plus 0.5em minus 0.4em\relax MDPI,
  2022, pp. 86--102.

\bibitem{fu_Collection_IoT21}
S.~Fu, Y.~Tang, Y.~Wu, N.~Zhang, H.~Gu, C.~Chen, and M.~Liu, ``Energy-efficient
  {UAV}-enabled data collection via wireless charging: A reinforcement learning
  approach,'' \emph{IEEE Internet of Things Journal}, vol.~8, no.~12, pp.
  10\,209--10\,219, 2021.

\bibitem{adachi_20_tvt}
H.~Huang, Y.~Yang, H.~Wang, Z.~Ding, H.~Sari, and F.~Adachi, ``Deep
  reinforcement learning for {UAV} navigation through massive {MIMO}
  technique,'' \emph{IEEE Transactions on Vehicular Technology}, vol.~69,
  no.~1, pp. 1117--1121, 2020.

\bibitem{Danhao22_cletter}
D.~Deng, C.~Wang, and W.~Wang, ``Joint air-to-ground scheduling in {UAV}-aided
  vehicular communication: A {DRL} approach with partial observations,''
  \emph{IEEE Communications Letters}, vol.~26, no.~7, pp. 1628--1632, 2022.

\bibitem{Flow_level19}
V.~Saxena, J.~Jaldén, and H.~Klessig, ``Optimal {UAV} base station
  trajectories using flow-level models for reinforcement learning,'' \emph{IEEE
  Transactions on Cognitive Communications and Networking}, vol.~5, no.~4, pp.
  1101--1112, 2019.

\bibitem{Shahriar_JCN20}
S.~A. Al-Ahmed, M.~Z. Shakir, and S.~A.~R. Zaidi, ``Optimal {3D UAV} base
  station placement by considering autonomous coverage hole detection, wireless
  backhaul and user demand,'' \emph{Journal of Communications and Networks},
  vol.~22, no.~6, pp. 467--475, 2020.

\bibitem{Yitong_iot22}
Y.~Liu, J.~Yan, and X.~Zhao, ``Deep-reinforcement-learning-based optimal
  transmission policies for opportunistic {UAV}-aided wireless sensor
  network,'' \emph{IEEE Internet of Things Journal}, vol.~9, no.~15, pp.
  13\,823--13\,836, 2022.

\bibitem{liu2018effects}
R.~Liu and J.~Zou, ``The effects of memory replay in reinforcement learning,''
  in \emph{2018 56th annual allerton conference on communication, control, and
  computing (Allerton)}.\hskip 1em plus 0.5em minus 0.4em\relax IEEE, 2018, pp.
  478--485.

\bibitem{sun2020adaptive}
Q.~Sun, W.-X. Zhou, and J.~Fan, ``Adaptive {Huber} regression,'' \emph{Journal
  of the American Statistical Association}, vol. 115, no. 529, pp. 254--265,
  2020.

\bibitem{Tiankui_safe_dqn21}
T.~Zhang, J.~Lei, Y.~Liu, C.~Feng, and A.~Nallanathan, ``Trajectory
  optimization for {UAV} emergency communication with limited user equipment
  energy: A safe-{DQN} approach,'' \emph{IEEE Transactions on Green
  Communications and Networking}, vol.~5, no.~3, pp. 1236--1247, 2021.

\bibitem{graves2013generating}
A.~Graves, ``Generating sequences with recurrent neural networks,'' \emph{arXiv
  preprint arXiv:1308.0850}, 2013.

\bibitem{goodfellow2016deep}
I.~Goodfellow, Y.~Bengio, and A.~Courville, \emph{Deep learning}.\hskip 1em
  plus 0.5em minus 0.4em\relax MIT press, 2016.

\bibitem{batch_norm_2015}
S.~Ioffe and C.~Szegedy, ``Batch normalization: Accelerating deep network
  training by reducing internal covariate shift,'' in \emph{International
  conference on machine learning}.\hskip 1em plus 0.5em minus 0.4em\relax pmlr,
  2015, pp. 448--456.

\bibitem{srivastava2014dropout}
N.~Srivastava, G.~E. Hinton, A.~Krizhevsky, I.~Sutskever, and R.~Salakhutdinov,
  ``Dropout: A simple way to prevent neural networks from overfitting,''
  \emph{Journal of Machine Learning Research}, vol.~15, no.~1, pp. 1929--1958,
  2014.

\bibitem{halim_ini}
F.~{Lagum}, I.~{Bor-Yaliniz}, and H.~{Yanikomeroglu}, ``Strategic densification
  with {UAV-BSs} in cellular networks,'' \emph{IEEE Wireless Commun. Lett.},
  vol.~7, no.~3, pp. 384--387, 2018.

\bibitem{wrc19-al}
{International Telecommunication Union}, ``{RESOLUTION 247 (WRC-19):
  Facilitating mobile connectivity in certain frequency bands below 2.7 GHz
  using high-altitude platform stations as International Mobile
  Telecommunications base stations},'' in \emph{The World Radiocommunication
  Conference}, 2019.

\end{thebibliography}

\end{document}